\documentstyle[12pt]{article}
\begin{document}

\hfill DUKE-CGTP-99-06

\hfill hep-th/9909108

\vspace{1.25in}

\begin{center}

{\large\bf Discrete Torsion and Gerbes I}

\vspace{0.75in}

Eric R. Sharpe \\
Department of Physics \\
Box 90305 \\
Duke University \\
Durham, NC  27708 \\
{\tt ersharpe@cgtp.duke.edu} \\

 $\,$

\end{center}

In this technical note we give a purely geometric understanding
of discrete torsion, as an analogue of orbifold Wilson lines for
two-form tensor field potentials.  
In order to introduce discrete torsion in this
context, we describe gerbes and the description of certain
type II supergravity
tensor field potentials as connections on gerbes.  Discrete torsion then
naturally appears in describing the action of the orbifold group
on (1-)gerbes, just as orbifold Wilson lines appear in describing
the action of the orbifold group on the gauge bundle.
Our results are not restricted to trivial gerbes -- in other words,
our description of discrete torsion applies equally well to
compactifications with nontrivial $H$-field strengths.
We also describe
a specific program for rigorously deriving analogues of discrete torsion for
many of the other type II tensor fields, and we are able to make specific
conjectures for the results.

\begin{flushleft}
September 1999
\end{flushleft}


\newpage

\tableofcontents

\newpage

\section{Introduction}

Historically discrete torsion has been a rather mysterious
aspect of string theory.  Discrete torsion was originally
discovered \cite{vafadt} as an ambiguity in the choice of
phases to assign to twisted sectors of string orbifold partition
functions.  Although other work has been done on the subject
(see, for example, \cite{vafaed,amg,edbaryons,md1,md2}), no work done to date
has succeeded in giving any sort of 
genuinely deep understanding
of discrete torsion.  In fact, discrete torsion has sometimes been
referred to has an inherently stringy degree of freedom, without
any geometric analogue.

In this paper we shall give a purely geometric understanding of discrete
torsion.  Specifically, we describe  
discrete torsion as a precise analogue of orbifold
Wilson lines, for 2-form fields rather than vector fields.
Put another way, we shall argue that discrete torsion can be
understood as ``orbifold Wilson surfaces.''

Our description of discrete torsion hinges on a deeper understanding
of type II $B$-fields than is common in the literature.
More specifically, just as vector potentials (gauge fields)
are understood as connections
on bundles, we describe $B$-fields as connections on (1-)gerbes.
Although gerbes seem to be well-known in some circles, their
usefulness does not seem to be widely appreciated.
We shall review a recent description of transition functions for
gerbes, given in \cite{hitchin,dcthesis}, which provides a simplified language
in which to discuss gerbes, and then shall discuss gerbes themselves
(in the language of stacks) in detail in \cite{dt2}.  
As accessible accounts of gerbes which provide
the level of detail we need do not seem to exist, we provide
such an overview in \cite{dt2}.  In a later paper \cite{dt3} we shall
provide a simplified way of understanding orbifold group actions on
$B$ fields, and shall also derive additional physical manifestations
of discrete torsion from the ideas presented here.

Let us take a moment to give some general explanation of our
ideas.  In defining an orbifold of a physical theory, 
the orbifold group $\Gamma$ must
define a symmetry of the theory.  Specifying the action of the
orbifold group on the underlying topological space is,
however, not sufficient when bundles or other objects (such
as gerbes) are present -- specifying an action of the orbifold
group $\Gamma$ on a space does not uniquely specify an action of
$\Gamma$ on a bundle.  Put another way, for any given
action of $\Gamma$ on the base space, there can be multiple
inequivalent actions of $\Gamma$ on a bundle or a gerbe.
This fact is usually glossed over in descriptions of orbifolds.

What is the physical meaning of this ambiguity in the lifting
of the action of $\Gamma$ to a bundle?
Specifying a specific action of $\Gamma$ on a bundle
with connection implicitly defines an orbifold Wilson line.
In other words, orbifold Wilson lines are precisely 
choices of actions of $\Gamma$ on a bundle with connection.
We shall show that, similarly, discrete torsion is a choice of
an action of $\Gamma$ on a (1-)gerbe with connection.

Technically, specifying a lift of $\Gamma$ to a bundle, or a gerbe,
is known as specifying an ``equivariant structure'' on the
bundle or gerbe.  Thus, in the paper we shall often speak
of classifying equivariant structures, which means classifying
lifts of $\Gamma$.

Our results are not restricted to trivial gerbes -- in other words,
our description of discrete torsion applies equally well to
compactifications with nontrivial $H$-field strengths.
Also, we do not make any assumptions concerning the nature of
$\Gamma$ -- it does not matter whether $\Gamma$ is abelian or nonabelian.
It also does not matter whether $\Gamma$ acts freely on the
underlying topological space -- in our description, freely-acting
orbifolds are understood in precisely the same way as orbifolds
with fixed points.

We are also able to describe analogues of discrete torsion for the
type IIA RR 1-form and the IIB RR 2-form fields.
In addition, our approach makes it clear that there
should exist analogues of discrete torsion for the other
tensor fields appearing in supergravity theories.  We describe
a specific program for rigorously deriving analogues of discrete torsion for
many of the other type II tensor field potentials
(specifically, those which can be understood in terms of gerbes), and
conjecture the results -- that analogues of discrete torsion
for $p$-form fields are measured by $H^p(\Gamma, U(1))$,
where $\Gamma$ is the orbifold group.

We begin in section~\ref{obl} by reviewing orbifold Wilson lines
in language that will easily generalize.  More specifically,
we describe orbifold Wilson lines as an ambiguity in lifting
the action of an orbifold group to a bundle with connection.
In section~\ref{ngerbe} we give a basic discussion of $n$-gerbes,
describing how they can be used to understand many of the tensor
field potentials appearing in supergravity theories.  This discussion is
necessary because we shall describe discrete torsion as a
precise analogue of orbifold $U(1)$ Wilson lines for 1-gerbes.
In section~\ref{dtmain} we outline precisely how one can
derive discrete torsion as an ambiguity in specifying the
action of an orbifold group on a 1-gerbe with connection,
in other words, as an analogue of orbifold Wilson lines.
We do not give a rigorous derivation of discrete torsion in this
paper; see instead \cite{dt2}.  Finally, we include an appendix
on group cohomology, which is used a great deal in this paper
and may not be familiar to the reader.

In this paper we concentrate on developing some degree of
physical intuition for our results on discrete torsion,
and give simplified (and rather loose) derivations.
A rigorous derivation of our results on 1-gerbes,
together with a detailed description of 1-gerbes, is
provided in a separate paper \cite{dt2}.

We should remark that our purpose in writing this paper and
\cite{dt2} was primarily conceptual, rather than computational,
in nature.  In these papers we give a new conceptual understanding
of discrete torsion.  Along the way we provide some fringe benefits,
such as an understanding of orbifold Wilson lines for nontrivial
bundles, a description of discrete torsion in backgrounds with
nontrivial torsion\footnote{Torsion in the sense of,
nontrivial 3-form curvature, as opposed to the mathematical senses.}, and
a thorough pedagogical review of 1-gerbes in terms
of sheaves of categories.  However, we do not provide significant
new computational methods.

We should mention that there are a few issues concerning discrete
torsion which we shall not address in either this paper or \cite{dt2}.
First, we shall only discuss discrete torsion for orbifold singularities.
One might wonder if discrete torsion, or some close analogue,
can be defined for non-orbifold singularities, such as conifold
singularities; we shall not address this matter here.
Second, we shall not attempt to discuss how turning on discrete
torsion alters the moduli space structure, i.e., how discrete torsion
changes the allowed resolutions of singularities.
See instead \cite{vafaed} for a preliminary discussion of this matter.
These matters will be discussed in \cite{dt3}.


We should also mention that in an earlier version of this
paper, we made a slightly stronger claim than appears here.
Namely, in earlier versions of this paper we claimed that the
difference between any two equivariant structures on a 1-gerbe
with connection is defined by an element of $H^2(\Gamma, U(1))$.
However, we have since corrected a minor error in \cite{dt2}.
Here, we only claim that the difference between equivariant structures
is an element of a group which includes $H^2(\Gamma, U(1))$ -- in other
words, there are additional degrees of freedom which we missed 
previously.  These additional degrees of freedom will be discussed
in more detail in \cite{dt3}.

\section{Orbifold Wilson lines}    \label{obl}

We shall begin by making a close examination of principal
bundles and heterotic string orbifolds, in order
to carefully review the notion of an ``orbifold Wilson line.''
In the next section, we describe gerbes, which provide
the generalization of line bundles required to describe
higher rank tensor potentials in supergravity theories.
Once we have given a basic description of gerbes,
we shall describe how the notion of orbifold Wilson line
generalizes to the case of gerbes, and in the process,
recover discrete torsion (and its analogues for the other
tensor potentials appearing in supergravity theories)
as a precise analogue of an orbifold Wilson line for a gerbe.

Let $X$ be a smooth manifold, and $\Gamma$ a discrete group acting 
by diffeomorphisms on $X$.  In this section we shall discuss
how to extend the action of $\Gamma$ to a bundle on $X$ and to
a connection on the bundle.  We shall explicitly recover,
for example, the orbifold Wilson lines that crop up
in toroidal heterotic orbifolds.  In the next section
we shall generalize the same methods to describe discrete torsion
as an analogue of orbifold Wilson lines for higher-degree gerbes.

\subsection{Basics of orbifold Wilson lines}   \label{basicobw}

Before we begin discussing orbifold Wilson lines in mathematical
detail, we shall take a moment to discuss them in
generality.  We should point out that in this paper we will
always assume that gauge bundles in question are abelian;
that is, any principal $G$-bundle appearing implicitly or
explicitly has abelian $G$.

In constructing heterotic orbifolds, people often
mention orbifold Wilson lines.  What are they?
In constructing a heterotic toroidal orbifold, one can 
combine the action
of $\Gamma$ on $X$ with a gauge transformation, so as
to create Wilson lines on the quotient space. 
The action on the gauge bundle defines the lift of $\Gamma$ to
the bundle. 
Such Wilson
lines are typically called orbifold Wilson lines.

The simple description above works precisely in the case that
the bundle being orbifolded is trivial.  In this case, 
there is a canonical (in fact, trivial) lift that leaves the fibers invariant,
and any other lift can be described as combining a gauge transformation
with the action of $\Gamma$ on the base.  In general, when the bundle
is nontrivial, there is no canonical lift, and so one
has to work harder.  A specification of a lift of $\Gamma$ 
to a bundle is
known technically as a choice of ``equivariant structure''
on the bundle,
and so to derive orbifold Wilson lines in the general context
we will speak of classifying equivariant structures.
We shall study equivariant structures in much more detail in the
next subsection.  In the remainder of this subsection, we shall attempt
to give some intuition for the relevant ideas.

Consider for simplicity the special case of an orbifold group
$\Gamma$ acting freely (without fixed points) on a space $X$.
How precisely do we describe a Wilson line on the quotient space?
Let $x \in X$, and pick some path
from $x$ to $g \cdot x$ for some $g \in \Gamma$.
In essence,
a Wilson loop on the quotient space $X/\Gamma$
is the composition of the
(nonclosed) Wilson loop along this path
from $x$ to $g \cdot x$ with a gauge transformation
describing the action of $g \in \Gamma$ on the corresponding
principal bundle.  It should be clear from this description that
equivariant structures on a bundle are encoding information about
Wilson lines on the quotient space, among other things.
For this reason, choices of equivariant structures are often
called orbifold Wilson lines. 


How should orbifold Wilson lines be classified?
Again, for simplicity assume $\Gamma$ acts freely on $X$.
We shall examine how flat connections on the quotient space
are related to flat connections on the cover, in order to 
shed some light.  (In later sections we shall not assume that bundles
under consideration
admit flat connections; we make this assumption here in order
to perform an enlightening calculation.)
First, recall that for any $G$, the moduli space of flat $G$-connections
on $X/\Gamma$, for abelian $G$, is given by
\begin{displaymath}
\mbox{Hom}\left( \pi_1(X/\Gamma), G \right) / G
\end{displaymath}
where $G$ acts by conjugation.  For abelian $G$,
conjugation acts trivially, and so the moduli space of
flat $G$-connections on $X/\Gamma$ is simply 
\begin{displaymath}
\mbox{Hom}\left( \pi_1(X/\Gamma), G \right)
\end{displaymath}
Thus, in order to study orbifold Wilson lines on $X$,
we need to understand how $\pi_1(X/\Gamma)$ is related to
$\pi_1(X)$.
Assuming $\Gamma$ is discrete and $X$ is connected,
then from the long exact
sequence for homotopy\footnote{Applied to the principal
$\Gamma$ bundle
\begin{displaymath}
\Gamma \: \longrightarrow \: X \: \longrightarrow \: X/\Gamma
\end{displaymath}
whose existence follows from the fact that $\Gamma$ acts freely.}
we find the short exact sequence
\begin{displaymath}
0 \: \longrightarrow \: \pi_1(X) \: \longrightarrow \:
\pi_1(X/\Gamma) \: \longrightarrow \: \pi_0(\Gamma) \:
\longrightarrow \: 0
\end{displaymath}
so $\pi_1(X/\Gamma)$ is an extension of $\pi_0(\Gamma) \cong
\Gamma$ by
$\pi_1(X)$.
As $\pi_1(X/\Gamma)$ receives a contribution from $\Gamma$,
we see that orbifolding enhances the space of possible Wilson lines 
by $\mbox{Hom}(\Gamma, G)$, roughly speaking.
More precisely, we have the long exact sequence
\begin{displaymath}
0 \: \longrightarrow \: \mbox{Hom}(\Gamma, G) \: \longrightarrow
\: \mbox{Hom}\left( \pi_1(X/\Gamma), G \right) \: \longrightarrow \:
\mbox{Hom}\left( \pi_1(X), G \right) \: \longrightarrow \: \cdots
\end{displaymath}
For example, for the special case $G = U(1)$, we have the
short exact sequence\footnote{Using the fact that $U(1) = {\bf R}/{\bf Z}$
is an injective ${\bf Z}$-module \cite[section I.7]{hs}.}
\begin{displaymath}
0 \: \longrightarrow \: H^1(\Gamma, U(1))  \:
\longrightarrow \: \mbox{Hom}\left( \pi_1(X/\Gamma), U(1) \right) \:
\longrightarrow \: \mbox{Hom}\left( \pi_1(X), U(1) \right) 
\: \longrightarrow \: 0
\end{displaymath}
where $H^1(\Gamma, U(1))$ denotes group cohomology of $\Gamma$
with trivial action on the coefficients $U(1)$.
Thus, we see explicitly that for $\Gamma$ discrete and freely-acting,
flat $U(1)$-connections on the quotient
space pick up a contribution from the group cohomology
group $H^1(\Gamma, U(1))$, which we can identify with
orbifold $U(1)$ Wilson lines.


The results of the discussion above are important and bear repeating.
We just argued that, for $\Gamma$ discrete and freely-acting,
flat $U(1)$ connections on the quotient get a contribution from 
$H^1(\Gamma, U(1))$.
We shall argue in later sections that for general abelian $G$
and general discrete $\Gamma$ (not necessarily freely-acting),
orbifold $G$ Wilson lines are classified by $H^1(\Gamma, G)$,
where $H^1(\Gamma, G)$ denotes group cohomology of $\Gamma$,
with coefficients\footnote{Technically, we are also assuming
that the action of $\Gamma$ on the coefficients $G$ is trivial.
We shall make this assumption on group cohomology throughout
this paper.} $G$.  In later sections we shall also
not make any assumptions concerning the nature of the
bundle -- we shall not assume the bundle in question
admits flat connections.  We shall rigorously derive the classification
of orbifold Wilson lines as a classification of equivariant structures
on principal bundles with connection.  When we classify equivariant
structures on gerbes with connection\footnote{And band $C^{\infty}(U(1))$,
technically.}, we shall recover a classification which includes
$H^2(\Gamma,U(1))$.

At this point we shall take a moment to clarify an issue that
may have been puzzling the reader.  We claimed in the introduction
that we would describe discrete torsion in terms of orbifold Wilson
lines for $B$ fields.  However, discrete torsion is measured in
terms of group cohomology, whereas (for flat connections)
Wilson lines are given by
$\mbox{Hom}(\pi_1, G)/G$.  However, for the special case $G = U(1)$,
\begin{displaymath}
\mbox{Hom}\left( \pi_1, G \right) / G \: = \: H^1(\Gamma, U(1))
\end{displaymath}
where $H^1(\Gamma, U(1))$ is group cohomology.  
It should now be clear to the reader that the usual
classification of discrete torsion -- given by
$H^2(\Gamma, U(1))$ -- is quite similar to this formal classification
of orbifold $U(1)$ Wilson lines -- given by $H^1(\Gamma, U(1))$.
In particular, the reader should now be less surprised that
orbifold Wilson lines and discrete torsion are related.


One issue we have glossed over so far
concerns ``fake'' Wilson lines, which we
shall now take a moment to discuss.  Consider for example the
orbifold ${\bf C}^2 / {\bf Z}_2$.  This space is simply-connected,
yet the usual prescriptions for orbifold Wilson lines tell us
that there is a physical degree of freedom (given by $\mbox{Hom}({\bf Z}_2, G)$)
which we would usually associate with Wilson lines.
Such degrees of freedom are often referred to as fake Wilson lines
\cite{dixonthesis}.

This degree of freedom is in fact physical -- not some
unphysical artifact.  In the next few sections we shall see
mathematically that one will recover degrees of freedom measured
by 
\begin{displaymath}
H^1(\Gamma, G) \: = \: \mbox{Hom}(\Gamma, G)
\end{displaymath}
for $\Gamma$-orbifolds of spaces with $G$-bundles (with $G$ abelian), 
regardless of
whether or not $\Gamma$ is freely acting.

How precisely should fake Wilson lines be interpreted on the
quotient space?  It can be shown \cite[chapter 14]{duistermaat} 
that if one quotients the total spaces of bundles, using equivariant
structures defining fake Wilson lines, then the resulting object
over the quotient space is not a fiber bundle.
(For example, a ${\bf Z}_2$ orbifold of a rank $n$ complex vector bundle
over ${\bf C}^2$ with nontrivial orbifold Wilson lines is not a fiber bundle
over the quotient space ${\bf C}^2/{\bf Z}_2$ -- one gets an object 
whose fiber over most points is ${\bf C}^n$, appropriate for a rank $n$
vector bundle, but whose fiber over the singularity is ${\bf C}^n/{\bf Z}_2$.)
We have not pursued this question in depth, but we do have
a strong suspicion.
In the case that $X/\Gamma$ is an algebraic variety\footnote{
Technically, a Noetherian normal variety.}, 
it is possible to construct (reflexive) sheaves which are closely
related to, but not quite the same as, bundles.  For example,
on ${\bf C}^2/{\bf Z}_2$, in addition to line bundles there
are also reflexive rank 1 sheaves.  We find it very tempting to
conjecture that these reflexive rank 1 sheaves correspond to
quotients of equivariant line bundles on ${\bf C}^2$ with nontrivial
fake Wilson lines, and that more generally, fake Wilson lines
on quotient spaces that are algebraic varieties have an interpretation
in terms of reflexive sheaves which are not locally free.
Moreover, isomorphism classes of
reflexive sheaves on affine spaces ${\bf C}^2/\Gamma$ are classified in the
same way as orbifold Wilson lines \cite{artinverdier}, a fact that forms
one corner of the celebrated McKay correspondence.
We shall have nothing further to say on this matter in this paper.

Before we move on to discuss lifts $\tilde{\Gamma}$ of $\Gamma$ acting
on line bundles with connection, we shall discuss some
amusing technical points regarding equivariant bundles.
One natural question to ask is the following:  given some
equivariant structure on a bundle $P$, how can one compute the characteristic
classes of the quotient bundle $P/\Gamma$?

The basic idea is to construct a principle $G$-bundle on
$E \Gamma \times_{\Gamma} X$, such that the projection to
$X/\Gamma$ yields the quotient bundle.  
We shall not work out the details here; see instead \cite{freedvafa},
where this program is pursued in detail.  In principle one could
follow the same program for the equivariant gerbes we shall
construct in later sections, and discuss their equivariant characteristic
classes.  However, we shall not pursue this direction in this paper.

In passing we should also note that on rare occasions,
equivariant bundles and equivariant bundles with connection
have been discussed in the physics literature in terms
of ``V-bundles'' \cite{gly,satake,kawasaki}.  The language
of V-bundles is rather different from the language we shall use
in this paper to describe
orbifold Wilson lines, though it is technically equivalent.

\subsection{Equivariant bundles}

Let $P$ be a principal $G$-bundle on $X$ for some abelian Lie
group $G$ (e.g., $G = U(1)^n$ for some positive $n$).
Given the action of $\Gamma$ on $X$, we would like to study lifts of
the action of $\Gamma$ on $X$ to the total space of $P$.

What precisely is a lift of the action of $\Gamma$?
Let $\pi: P \rightarrow X$ denote the projection, then a lift
of the action of an element $g \in \Gamma$ is a diffeomorphism $\tilde{g}:
P \rightarrow P$ such that $\tilde{g}$ is a morphism
of principal $G$-bundles.  The statement that 
$\tilde{g}$ is a morphism of bundles means precisely that
the following diagram commutes:
\begin{equation}  \label{liftdefn}
\begin{array}{ccc}
P & \stackrel{ \tilde{g} }{ \longrightarrow } & P \\
\makebox[0pt][r]{ $\scriptstyle{ \pi }$ } \downarrow & & 
\downarrow \makebox[0pt][l]{ $\scriptstyle{ \pi }$ } \\
X & \stackrel{ g }{ \longrightarrow } & X
\end{array}
\end{equation}
The statement that $\tilde{g}$ is a morphism of principal $G$-bundles,
not merely a morphism of bundles,
means that, in addition to the commutativity of the diagram above,
the action of $G$ on the total space must commute with $\tilde{g}$.
In other words, $\tilde{g}( h \cdot p) = h \cdot \tilde{g}(p)$ for
all $h \in G$ and $p \in P$.
Furthermore, in order to lift the action of the entire
group $\Gamma$ and not just its elements,
we shall impose the constraint that
\begin{equation}   \label{liftconstraint}
\widetilde{g_1 \cdot g_2} \: = \: \tilde{g}_1 \circ \tilde{g}_2
\end{equation}
for all $g_1, g_2 \in \Gamma$.

The constraint given in equation~(\ref{liftconstraint}) is quite
important; one is not always guaranteed of finding lifts that
satisfy~(\ref{liftconstraint}).  As an example\footnote{We would like
to thank A.~Knutson for pointing out this example to us.}, we shall examine the
nontrivial ${\bf Z}_2$
bundle over $S^1$.  Consider the group $\Gamma = {\bf Z}_2$
that acts on the base $S^1$ as a half-rotation of the circle.
On the total space of the nontrivial ${\bf Z}_2$ bundle,
essentially a $720^{\circ}$ object, $\Gamma$ must act by
rotation by either $+180^{\circ}$ or $-180^{\circ}$, in order
to cover the action of ${\Gamma}$ on the base $S^1$
(i.e., in order for diagram~(\ref{liftdefn}) to commute).
Unfortunately, neither such action on the total space of the bundle
squares to the identity, and so equation~(\ref{liftconstraint})
can not be satisfied in this case.

As the example just given demonstrates, although trivial bundles 
admit lifts of
orbifold group actions, not all nontrivial bundles admit
lifts of orbifold group actions.
Rather than digress to explain conditions for the existence of
lifts, we shall simply assume lifts exist in all examples considered
in this paper.  (We shall make a similar assumption when discussing 
equivariant gerbes.)

A lift of the action of $\Gamma$ to a bundle is also called
a choice of ($\Gamma$-)equivariant structure on the bundle.

In passing, we should also mention that instead of speaking of
lifts, we could equivalently work in terms of pullbacks.
Loosely speaking, in terms of pullbacks, 
a bundle $P$ is ``almost equivariantizable'' with respect to the action
of $\Gamma$ if, for all $g \in \Gamma$, $g^* P \cong P$.
As above, not all bundles will necessarily be equivariant with
respect to some given $\Gamma$, but we shall not discuss
relevant constraints in this paper.
More precisely, an equivariant bundle $P$ is defined to be 
a bundle with  
a choice of equivariant structure, which can be defined as
a specific set
of isomorphisms $\psi_g: g^* P \stackrel{ \sim }{ \longrightarrow }
P$ for all $g \in \Gamma$, subject to the obvious analogue 
of equation~(\ref{liftconstraint}).

It is easy to check that the definitions of equivariant structures
in terms of lifts and in terms of pullbacks 
are equivalent.  For completeness, we shall
outline the argument here.  Let $\psi_g: g^* P \stackrel{ \sim }{
\longrightarrow } P$ define an equivariant structure (in terms of
pullbacks) on a
principal bundle $P$.  Then, we can define a lift $\tilde{g}$
of $g \in \Gamma$ as, $\tilde{g} = \psi_{g^{-1}} \circ (g^{-1})^*$.
The reader can easily check that $\tilde{g}$ does indeed define
a lift of $g$, as defined above.  Conversely, given a lift
$\tilde{g}$ of $g \in \Gamma$, we can define an equivariant structure
(in terms of pullbacks)
$\psi_{g}: g^* P \stackrel{ \sim }{ \longrightarrow } P$
by, $\psi_g^{-1} = g^* \circ \tilde{g}$.  It is easy to check
that the two constructions given here are inverses of one another.

How many possible lifts of the action of a given $g \in \Gamma$
exist?  Given any one lift, we can certainly make any other
lift by composing the action of the lift with a gauge
transformation.  More precisely, given a set of lifts $\{ \tilde{g} | g \in 
\Gamma \}$,
we can construct a new set of lifts $\{ \tilde{g}' \}$ by composing each
$\tilde{g}$ with a gauge transformation $\phi_g: X \rightarrow G$
such that $\phi_{g_2}(x) \cdot \phi_{g_1}(g_2^{-1}x) = \phi_{g_2 g_1}(x)$
for all $x \in X$.
Moreover, any two lifts differ by a set of such gauge transformations.
We can rephrase this by saying that any two lifts of the action
of $\Gamma$ to $P$ differ by an element of
$\mbox{Hom}( \Gamma, C^{\infty}(G) )
 = H^1( \Gamma, C^{\infty}(G))$.


Now, from our knowledge of orbifold Wilson lines,
we will eventually want (equivalence classes of) lifts to
be classified by $H^1(\Gamma, G) = \mbox{Hom}( \Gamma, G)$,
but above we only have $H^1(\Gamma, C^{\infty}(G) )$.
What have we forgotten?

So far we have only studied how to extend the action of $\Gamma$
to the total space of a bundle.  We have not yet spoken about
further requiring the action of $\Gamma$ to preserve the connection
on the bundle.  This requirement will place additional constraints
on the lifts.
When we are done, we will see that by considering lifts
of the action of $\Gamma$ to line bundles with connection,
instead of just line bundles, we will recover the
classification $H^1(\Gamma, G)$, as desired.

For more information on equivariant bundles, see for example
\cite{duistermaat}.

\subsection{Equivariant bundles with connection}

In the previous subsection we described 
the action of the group $\Gamma$ on principal $G$-bundles,
for $G$ an abelian Lie group.  In this section we shall
extend this discussion to include consideration of a connection
on a bundle.  We shall argue that equivalence classes
of lifts of the action of $\Gamma$ to pairs (line bundle,
connection) are classified by the group cohomology
group $H^1( \Gamma, G )$.  (More precisely,
we shall find a non-canonical correspondence between equivariant
structures on principal $G$-bundles with connection and
elements of $H^1(\Gamma, G)$.  In special cases, such as 
trivial principal $G$-bundles, there will be a canonical
correspondence.)  For more information on
connection-preserving lifts, see\footnote{There is also
related information in \cite[sections 2.4, 2.5]{brylinski} and
\cite[section V.2]{gs}.  These references analyze a distinct but
related problem; their discussion might at first confuse the reader.
Specifically, instead of considering representations of $\Gamma$
in the group of all connection-preserving lifts of diffeomorphisms
of the base, they study the space of connection-preserving lifts
itself, and argue that it is a central extension of 
the group of bundle-with-connection-preserving
diffeomorphisms of the base by $U(1)$, for principal $U(1)$-bundles.
The reader might be then tempted to argue that lifts of $\Gamma$
should be classified by $H^2(\Gamma, U(1))$, but this is not quite
correct.  In particular, when viewing the set of all connection-preserving
lifts as a central extension, the elements that project to $\Gamma$
will not, in general, form a representation of $\Gamma$, i.e.,
will not satisfy equation~(\ref{liftconstraint}).
} \cite{freedvafa} and \cite[section 1.13]{kostant}.

Before going on, we shall take a moment to very briefly review connections
on principal bundles
and what it means for a lift to preserve a connection.
One way to think of a connection on a bundle is as a 
set of gauge fields $A_{\mu}$, one for each element of a good
cover of the base.  However, there is a slightly more elegant
description which we shall use instead \cite[section Vbis.A]{amp1}.  
If $P$ is the total
space of a principal $G$-bundle on $X$, then a connection
on $P$ is a map $TP \rightarrow \mbox{Lie }G$, or a 
$(\mbox{Lie }G)$-valued 1-form on $P$, satisfying certain
properties we shall not describe here (see instead,
for example, \cite[section 11.4]{martin} or
\cite[section V bis A1]{amp1}).  Given an open set $U \subset X$
such that $P|U$ is trivial, let $s: U \rightarrow P$ be a section,
and let $\alpha$ denote the connection on $P$, i.e., the 
corresponding $(\mbox{Lie }G)$-valued
1-form,
then we can recover a gauge field on the base 
$X$ as $s^* \alpha$.  Any two distinct sections $s_1, s_2: U \rightarrow P$
define gauge fields differing by a gauge transformation, i.e.,
$s_1^* \alpha = s_2^* \alpha - (d g) g^{-1}$.
If $\phi: X \rightarrow G$ defines a gauge transformation,  
then it acts on the connection $\alpha$ as 
(\cite[section V bis, problem 1]{amp1},
\cite[section 1.10]{kostant})
\begin{displaymath}
\alpha(p) \: \rightarrow \: \phi( \pi(p)) \, \alpha(p) \, \phi^{-1}( \pi(p))
\: - \: d \ln ( \phi \circ \pi ) (p)
\end{displaymath}
for $p \in P$ and $\pi: P \rightarrow X$ the projection.
Clearly, a gauge transformation $\phi: X \rightarrow G$ will preserve
the connection (not just up to gauge equivalence) if and only if
$\phi$ is a constant map.

How does a morphism of principal bundles act on a connection?
Let $\tau: P_1 \rightarrow P_2$ denote a morphism of principal bundles,
then if $\alpha_2$ is a connection on $P_2$, $\tau^* \alpha_2$
(defined in the obvious way) is a connection on $P_1$.
More relevantly to the problem under discussion, if $g \in \Gamma$
and $\tilde{g}$ denotes the lift of $g$, then we shall say
that $\tilde{g}$ preserves the connection $\alpha$ if
$\tilde{g}^* \alpha = \alpha$, not just up to gauge
transformation.


In order to get a well-defined connection living over the quotient $X/\Gamma$,
we shall demand that the lift of the action of $\Gamma$
preserves the connection itself, not
just its gauge-equivalence
class.  (If this were not the case, then we would not be able
to immediately write down a well-defined connection over the quotient
space.)
Phrased another way, a lift of the action of $\Gamma$ on $X$ will
yield a well-defined connection on the quotient precisely if it
can be deformed by an element of $H^1(\Gamma, C^{\infty}(G))$ so that
it preserves the connection itself, not just its gauge-equivalence
class.  Phrased another way still,
if we merely demanded that the lift of $\Gamma$ preserve
only the gauge-equivalence class of the connection, then naively
spoke of the gauge-equivalence class descending to the quotient,
we would not be guaranteed of finding any representatives of the
quotiented gauge-equivalence class.

Necessary and sufficient conditions for a lift of $\Gamma$ to preserve
the connection are known and easy to describe \cite{kostant}.
In fact, the action of $g \in \Gamma$ on the base $X$ is liftable
to a map of bundle with connection if and only if the action of
$g$ preserves the values\footnote{Strictly speaking,
preserves the values of the Wilson loops up to conjugation; however,
for bundles with abelian structure group, conjugation acts
trivially.} of Wilson loops on the base 
\cite[prop. 1.12.2]{kostant}.  (Note that even for bundles with a non-flat
connection -- nontrivial bundles -- one can still define Wilson 
loops -- however, only in the 
special case of a flat connection will the value of a Wilson loop
depend only on the homotopy class of the loop.)  
The reader should not be surprised
by this result, as this fact is often implicitly claimed in the literature
on toroidal heterotic orbifolds, for example.

Now, how many lifts of $\Gamma$ preserve the connection itself?
Let $\{ \tilde{g} \}$ denote a lift of $\{ g \in \Gamma \}$ that preserves
the connection itself.  We can compose $\{ \tilde{g} \}$ with
an element of $H^1( \Gamma, C^{\infty}(G) )$ to get another lift,
but only the constant elements, namely those in $H^1(\Gamma,
G) \subset H^1(\Gamma, C^{\infty}(G))$, will act so as to preserve
the connection itself.  Thus, $H^1(\Gamma, G)$ acts 
on the set of connection-preserving lifts of $\Gamma$, and it should
be clear this action is both transitive and free.

Note that in the very special case that the equivariant bundle
on $X$ is trivial, then there is (distinguished)
trivial lift, and so there is a canonical correspondence
between elements of $H^1(\Gamma, G)$ and connection-preserving
lifts.  
For more general bundles, there is no such distinguished lift.

As this result is important, we shall repeat it.
We have just shown that ($\Gamma$-)equivariant structures on $G$-bundles
with connection can be (noncanonically) identified with
elements of $H^1(\Gamma, G)$.  In special cases, such as trivial
bundles, there is a canonical identification.

In passing, note that our analysis of equivariant structures
on bundles with connection did not assume $\Gamma$ was freely-acting
or that $\Gamma$ be abelian:  our analysis applies equally
well to cases in which $\Gamma$ has fixed points on the base space,
as well as cases in which $\Gamma$ is nonabelian.

\subsection{Example:  heterotic orbifolds}

As a more explicit example, let us consider how to define a toroidal
heterotic orbifold.  Here we have some principal $G$ bundle
(for some $G$)
over the torus, which for simplicity we shall assume to
be trivial\footnote{Although the bundle has a flat connection,
it need not be topologically trivial or even trivializable -- this
is a stronger constraint than necessary, which we are 
introducing in order to keep this warm-up example simple.}.
We shall also assume the connection on the bundle on the torus
is not merely flat, but actually trivial.  In these special circumstances,
any diffeomorphism of the base lifts to an action on the bundle.

Now, to define a lift of an action of $\Gamma$ on the torus to the
total space of the principal bundle is trivial.  Since the total
space of a trivial principal bundle is just $X \times G$, clearly
we can lift the action of $\Gamma$ to the total space by defining it
to be trivial on the fiber $G$.  (More generally, for a nontrivial bundle,
demanding that the group $\Gamma$ lift to an action on the total space
of the bundle is not trivial.  Depending upon both $X$ and the
bundle in question, there are often obstructions.)
Given any one such lift, we can find all other possible
lifts simply by composing the trivial lift with a gauge transformation.

In order to get a well-defined connection on the quotient space,
however, there are some constraints on the possible lifts.
First, note that in these special circumstances, we can
describe any lift as the composition of the trivial lift with
a gauge transformation.  For any $g \in \Gamma$, let
$\phi_g:  X \rightarrow G$ denote the corresponding gauge transformation.
Then in order to preserve the connection itself, $\phi_g$ must be
constant, in other words, $\phi_g = \epsilon(g)$ for some
$\epsilon: \Gamma \rightarrow G$.  These $\epsilon(g)$ define the usual
orbifold Wilson lines.

\subsection{Discussion in terms of \v{C}ech cohomology}

Eventually in this paper we will work through arguments closely
analogous to those above to derive analogues of orbifold Wilson
lines for gerbes.  In order to do this properly is somewhat
difficult and time-consuming -- (1-)gerbes are properly described
in terms of sheaves of categories, and their full analysis 
can be somewhat lengthy.  In order to give some general intuition
for the results at the level of Hitchin's \cite{hitchin,dcthesis}
discussion of gerbes, we will eventually give a rather loose
derivation of the results in terms of \v{C}ech cohomology.
(A rigorous derivation will appear in \cite{dt2}.)

As a warm-up for that eventual discussion, in this section we shall
very briefly describe how one can re-derive orbifold
Wilson lines while working at the level of \v{C}ech cohomology, i.e.,
at the level of transition functions for bundles.  
We feel that such an approach
is philosophically somewhat flawed -- the transition 
functions of a bundle do not really
capture all information about the bundle.
For example, gauge transformations
of a bundle are completely invisible at the level of transition functions.
Thus, we do not find the notion of defining an equivariant structure
on a bundle by putting an equivariant structure on its
transition functions to be completely satisfying.
Thus, when we study equivariant gerbes, we shall not limit ourselves
to only discussing equivariant structures on gerbe transition functions,
but shall also discuss equivariant structures on the gerbes themselves.

Experts will note that in this subsection we implicitly make
some assumptions regarding the behavior of bundle trivializations
under the action of the orbifold group.  As our purpose in this
subsection is not to give a
rigorous derivation but merely to perform an enlightening calculation,
we shall gloss over such issues.

Let $P$ be a principal $G$-bundle on a manifold $X$, and let 
$\Gamma$ be a group acting on $X$ by diffeomorphisms.
Let $\{ U_{\alpha} \}$ be a ``good invariant'' cover of $X$,
by which we means that each $U_{\alpha}$ is invariant under $\Gamma$,
and each $U_{\alpha}$ is a disjoint union of contractible open sets.
For example, one can often get good invariant covers of a space $X$
from good covers of the quotient $X / \Gamma$.  Note that a good invariant
cover is not a good cover, in general.  We shall assume good invariant
covers exist in this subsection, though it is not clear that this need
be true in general.  (Again, our purpose in this subsection is to
present enlightening calculations and plausibility arguments,
not completely rigorous proofs.)

Let $h_{\alpha \beta}$ denote transition functions for the bundle $P$.
Assume $P$ has an equivariant structure, 
which at the level of transition functions
means that for all $g \in \Gamma$ there exist functions $\nu^g_{\alpha}:
U_{\alpha} \rightarrow G$ such that
\begin{equation}   \label{hequiv1}
g^* h_{\alpha \beta} \: (\, = \, h_{\alpha \beta} \circ g \, ) \: = \:
\nu^g_{\alpha} \, h_{\alpha \beta} \, (\nu^g_{\beta})^{-1}
\end{equation}
The functions $\nu^g_{\alpha}$ are local trivialization realizations
of an isomorphism of principal $G$-bundles $\nu^g: P \stackrel{\sim}
{\longrightarrow} g^* P$ for each $g \in \Gamma$.
It should be clear that $\nu^g = (\psi_g)^{-1}$ where the
$\psi_g$ were defined in the section on equivariant bundles.

The $\nu^g$ partially specify an equivariant structure on $P$,
but we also need a little more information.  In particular, 
we must also demand that for $g_1, g_2 \in \Gamma$,
\begin{equation}   \label{nuequiv1}
\nu^{g_2}_{\alpha} \, g_2^* \nu^{g_1}_{\alpha} \: = \: \nu^{g_1 g_2}_{\alpha}
\end{equation}
Note that this is the appropriate \v{C}ech version of
equation~(\ref{liftconstraint}).

Now, suppose $\nu^g_{\alpha}$ and $\overline{\nu}^g_{\alpha}$ define
two distinct equivariant structures on $P$, with respect to the same
group $\Gamma$.  Define $\phi^g_{\alpha}: U_{\alpha} \rightarrow U(1)$ by,
\begin{displaymath}
\phi^g_{\alpha} \: \equiv \: \frac{ \nu^g_{\alpha} }{ \overline{\nu}^g_{\alpha}
}
\end{displaymath}
From the fact that both $\nu^g_{\alpha}$ and $\overline{\nu}^g_{\alpha}$
must satisfy equation~(\ref{hequiv1}), we can immediately derive 
the fact that
\begin{displaymath}
\phi^g_{\alpha} \: = \: \phi^g_{\beta}
\end{displaymath}
on $U_{\alpha \beta} = U_{\alpha} \cap U_{\beta}$, and so the $\phi^g_{\alpha}$
define a function $\phi^g: X \rightarrow U(1)$.
This is a gauge transformation describing the difference
between two equivariant structures.  It is almost, but not quite,
the same as the gauge transformation $\phi_g$ described in the
section on equivariant bundles.

From equation~(\ref{nuequiv1}), we see that the $\phi^g$ must obey
\begin{displaymath}
\phi^{g_2} \cdot g_2^* \phi^{g_1} \: = \: \phi^{g_1 g_2}
\end{displaymath}

Thus, any two equivariant structures on $P$ differ by an element
of $H^1(\Gamma, C^{\infty}(G))$, as described in the section on
equivariant bundles.

We shall now recover the fact
that equivariant structures on bundles with connection differ by
elements of $H^1(\Gamma, G)$, for abelian $G$. 

Let $A^{\alpha}$ be a ($\mbox{Lie }G$)-valued one-form on the open
set $U_{\alpha}$, defining a connection on $P$.
In other words, on overlaps $U_{\alpha \beta} = U_{\alpha} \cap
U_{\beta}$,
\begin{displaymath}
A^{\alpha} \, - \, A^{\beta} \: = \: d \, \mbox{ln } h_{\alpha \beta}
\end{displaymath}

Under the action of $g \in \Gamma$, since
\begin{displaymath}
g^* h_{\alpha \beta} \: = \: \nu^g_{\alpha} \, h_{\alpha \beta} \, 
(\nu^g_{\beta} )^{-1}
\end{displaymath}
we know that
\begin{displaymath}
g^* A^{\alpha} \: = \: A^{\alpha} \: + \: d \, \mbox{ln } \nu^g_{\alpha}
\end{displaymath}

From this we see that if $\nu^g_{\alpha}$ and $\overline{\nu}^g_{\alpha}$
define two distinct equivariant structures on the transition functions,
then we must have $\nu^g_{\alpha} / \overline{\nu}^g_{\alpha}$
be constant, in order for $g^* A^{\alpha}$ to be well-defined.
Thus, we recover the fact that $\phi^g$ is constant,
and so the subset of $H^1(\Gamma, C^{\infty}(G))$ that describes
equivariant bundles with connection is given by
$H^1(\Gamma, G)$, as claimed.

In essence, we have been using a form of equivariant \v{C}ech cohomology.
The mathematics literature seems to contain multiple\footnote{One
version of equivariant \v{C}ech cohomology is described in 
\cite[chapitre V]{grothcech}.  Another version is described
in
\cite[section 2]{jrw} and \cite[section 5]{rw}.} versions of
equivariant \v{C}ech cohomology, unfortunately none of them are quite adequate
for our eventual needs (i.e., none of them correspond to
the precise way we set up equivariant structures on gerbes), 
and so we shall not speak about them further.

\section{$n$-Gerbes}     \label{ngerbe}

Discrete torsion has long been associated with the two-form
$B$-fields of supergravity theories.  The reader should not
be surprised, therefore, that a deep understanding of discrete
torsion hinges on a deep understanding of the two-form $B$ fields.
We shall argue that $B$ fields should be understood as connections
on 1-gerbes, and that discrete torsion arises when lifting
the action of an orbifold group to a 1-gerbe with connection.

Why might one want a new mathematical object to describe
$B$-fields in type II string theory?
The reason is a dilemma that has no doubt puzzled for many
years.  The torsion\footnote{No relation to the various mathematical
concepts of torsion.} $H$ is defined to be $H = dB$.
So long as $H$ is taken to be cohomologically trivial, this is
certainly a sensible definition.  Unfortunately, in general
one often wants to speak of $H$ which is not a cohomologically
trivial element of $H^3({\bf R})$.  
In such a case, the relation $H = dB$ can only hold locally.
(This point has been made previously in, for example,
\cite{freedetal}.)

We shall see shortly that such $H$ can be understood globally
as a connection on a $1$-gerbe.  More generally, many of the tensor
field potentials appearing in type II string theories naively appear to
have a natural
and obvious interpretation in terms of connections on $n$-gerbes,
though for the sake of simplifying the discussion we will
usually only discuss the two-form tensor field in examples.

In passing, we should also mention that although some tensor field
potentials may be understood in terms of gerbes, it is not
clear that all tensor field potentials have such an understanding.
One notable exception is the $B$-field of heterotic
string theory.  Recall that one has the anomaly cancellation constraint
\begin{displaymath}
d H \: \propto \: \mbox{Tr } R \wedge R \: - \: \mbox{Tr } F \wedge F
\end{displaymath}
If the heterotic $B$ field were a 1-gerbe connection, 
then we shall see that the curvature $H$
would be a closed form, whereas that is certainly not the case here
in general.  Moreover, many other tensor field potentials have
nontrivial interactions and ``mixed'' gauge transformations, and it
is not at all clear whether these phenomena can be
understood in terms of gerbes.
As a result, one should be somewhat careful about blindly identifying
all tensor field potentials with connections on gerbes.
These issues should not arise for the comparatively simpler
cases of type II 2-form potentials (with other background
fields turned off), which is the primary case
of interest for us in this paper.

We should also mention a slight technical caveat.
We shall only be discussing gerbes with ``band'' $C^{\infty}(U(1))$
\cite{dt2}, which means, in less technical language, that there
exist more general gerbes than those discussed in this section.
For example, some theories contain multiple coupled tensor multiplets
(for one example, see \cite{ori1}), which would be described
in terms of connections on gerbes with ``band'' $C^{\infty}(U(1)^N)$.
We shall not discuss gerbes with general bands in this paper;
see instead \cite{dt2}.

In this section we will give a description of gerbes
and connections on gerbes, due to \cite{hitchin,dcthesis}.
We shall begin by defining gerbes themselves, then afterwards
we shall describe connections on gerbes.
In the next section we will discuss the analogue of
``orbifold Wilson lines'' for gerbes.

\subsection{Description in terms of cohomology}

We shall begin by describing characteristic classes
of abstract objects called ``$n$-gerbes,'' following
the discussion in \cite{hitchin,dcthesis}.
These characteristic classes, which for $n$-gerbes on a space
$X$, 
are
elements of the sheaf cohomology group
$H^{n+1}(X, C^{\infty}(U(1)) )$.  
This is closely analogous
to describing a line bundle in terms of Chern classes.
More intrinsic definitions of gerbes are given in the next
section and in \cite{dt2}.
Gerbes themselves take considerably longer to define;
by first describing their characteristic classes, we hope to give
the reader some basic intuitions for these objects.

In passing we should comment on our usage of the terminology
``$n$-gerbe.''  We are following the simplified conventions
of \cite{hitchin,dcthesis}.  In general, an $n$-gerbe should,
morally, be understood in terms of sheaves of multicategories.
Unfortunately, $n$-categories for $n > 2$ are not well
understood at present.  As a result, although $1$-gerbes
and, to a slightly lesser extent, $2$-gerbes are well understood,
higher degree gerbes are not on as firm a footing.
It seems reasonably clear that such objects should exist, however,
and one can certainly describe many properties that a general
$n$-gerbe should possess in terms of characteristic classes
(as in this section) and Deligne cohomology. 
Thus, we shall often speak (loosely) of general $n$-gerbes,
though for $n > 2$ the reader should probably take such
remarks with a small grain of salt.

A couple of paragraphs ago we mentioned that the characteristic
classes of gerbes could be understood in terms of 
sheaf cohomology, and more specifically that the characteristic classes
of possible $n$-gerbes on a space $X$ live in
$H^{n+1}(X, C^{\infty}(U(1)) )$.
For those readers not acquainted with sheaf cohomology,
we can express this somewhat more simply (and loosely) in terms
of \v{C}ech cocycles with respect to some fixed open cover.  Let $U_{\alpha}$
be a ``reasonably nice\footnote{Specifically, a good open cover.}'' open 
cover of $X$.  Then an element of 
$H^{n+1}(X, C^{\infty}( U(1) ) )$ is 
essentially defined by a set of smooth functions
$h_{\alpha_0 \cdots \alpha_{n+1} }: U_{\alpha_0 \cdots
\alpha_{n+1}} \rightarrow U(1)$, one for each 
overlap $U_{\alpha_0 \cdots \alpha_{n+1} } = 
U_{\alpha_0} \cap \cdots \cap U_{\alpha_{n+1} }$,
subject to the constraint
\begin{equation}  \label{cechclosed}
( \delta h )_{\alpha_0 \cdots \alpha_{n+2} } \: = \: 1
\end{equation}
where $\delta h$ is defined by
\begin{displaymath}
( \delta h )_{\alpha_0 \cdots \alpha_{n+2} } \: \equiv \:
\prod_{i=0}^{n+2} h^{(-)^{i}}_{\alpha_0 \cdots \hat{ \alpha_i } \cdots
\alpha_{n+2} } \\
\end{displaymath}
on the intersection $U_{\alpha_0} \cap \cdots \cap U_{\alpha_{n+2} }$.
Two such sets of functions $h_{\alpha_0 \cdots \alpha_{n+1}}$,
$h'_{\alpha_0 \cdots \alpha_{n+1}}$ are identified with the
same element of $H^{n+2}(X, C^{\infty}( U(1) ) )$ if and only if
\begin{equation}  \label{modexact}
h_{\alpha_0 \cdots \alpha_{n+1}} \: = \:
h'_{\alpha_0 \cdots \alpha_{n+1}} \,
\prod_{i=0}^{n+1} f^{(-)^{i}}_{\alpha_0 \cdots \hat{ \alpha_i }
\cdots \alpha_{n+1}}
\end{equation}
for some functions $f_{\alpha_0 \cdots \alpha_{n}}:
U_{\alpha_0} \cap \cdots \cap U_{\alpha_{n}} \rightarrow U(1)$.

As a special case, let us see how this duplicates line bundles.
In the classification of $n$-gerbes implicit in the description
of characteristic classes above, it should be clear that
line bundles are very special examples of $n$-gerbes,
specifically, $0$-gerbes.
A $0$-gerbe is
specified by an element of $H^1(X, C^{\infty}( U(1) ))$,
that is, a set of smooth functions $h_{\alpha \beta}:
U_{\alpha} \cap U_{\beta} \rightarrow U(1)$, such that
\begin{displaymath}
h_{\beta \gamma} h^{-1}_{\alpha \gamma} h_{\alpha \beta} \: = \: 1
\end{displaymath}
on triple intersections $U_{\alpha} \cap U_{\beta} \cap U_{\gamma}$
(this is the specialization of equation~(\ref{cechclosed})).
The reader should immediately recognize this as defining
transition functions for a smooth $U(1)$ bundle on $X$.
Equation~(\ref{cechclosed}) is precisely the statement that
transition functions agree on triple overlaps.
Moreover, two $U(1)$ line bundles are equivalent if and only
if their transition functions $h_{\alpha \beta}$, $h'_{\alpha
\beta}$ are related by \cite[chapter 5.2]{husemoller}
\begin{displaymath}
h_{\alpha \beta} \: = \: h'_{\alpha \beta} f_{\alpha} / f_{\beta}
\end{displaymath}
for some set of functions $f_{\alpha}: U_{\alpha} \rightarrow U(1)$.
The reader should immediately recognize this as the specialization
of equation~(\ref{modexact}).

Although the sheaf cohomology group $H^1(X, C^{\infty}(U(1)))$
precisely describes (equivalence classes of) transition functions
for $0$-gerbes (smooth principal $U(1)$ bundles),
the same is not true for higher degree gerbes -- 
an element of sheaf cohomology for a higher degree gerbe
does not define a set of transition functions. 
(We shall study transition functions for gerbes in the next
subsection.)

We can rewrite these characteristic classes of $n$-gerbes in a format
that is more accessible to calculation.
Using the short exact sequence of ($C^{\infty}$)
sheaves
\begin{displaymath}
0 \rightarrow C^{\infty}({\bf Z}) \cong {\bf Z} \rightarrow
C^{\infty}( {\bf R} ) \rightarrow C^{\infty}( U(1) )
\rightarrow 0
\end{displaymath}
one can immediately prove, from the associated long exact sequence,
that for $n \geq 0$,
\begin{displaymath}
H^{n+1}( X, C^{\infty}( U(1) ) ) \: \cong \: H^{n+2}( X, {\bf Z} )
\end{displaymath}

As a special case, this implies that $0$-gerbes are classified by
elements of $H^2(X, {\bf Z})$, and indeed it is a standard
fact that $C^{\infty}$ line bundles are classified by
their first Chern class.

In general, any two trivializations of a trivializable $n$-gerbe,
that is, one described by a cohomologically trivial $(n+1)$-cocycle,
differ by an $(n-1)$-gerbe.  This should be clear from the description
above -- any cohomologically trivial $(n+1)$-cocycle is a coboundary
of some $n$-cochain, and any two such cochains differ by an $n$-cocycle,
defining an $(n-1)$-gerbe.

Before going on, we should mention that in the remainder
of this paper (as well as \cite{dt2}) we shall usually abbreviate
``1-gerbes'' to simply ``gerbes.''  Unfortunately, on rare occasion 
we shall also use ``gerbes'' as shorthand for $n$-gerbes.
The usage should be clear from context.

\subsection{Description in terms of transition functions}

In the previous section we described $n$-gerbes in terms of sheaf cohomology,
which is precisely analogous to describing line bundles in terms of
Chern classes.  Traditionally gerbes are typically defined in terms of
sheaves of multicategories, as we shall do
in \cite{dt2}.  In this section, we shall
give a simplified account, due to \cite{hitchin,dcthesis}, which
amounts to describing gerbes in terms of transition functions.
In \cite{dt2} we shall review sheaves of categories and the
description of 1-gerbes in such language, in addition
to giving a geometric first-principles derivation of discrete torsion.

Before grappling with transition functions for $n$-gerbes, we shall
begin with a description of transition functions for 1-gerbes.
Let $\{ U_{\alpha} \}$ be a good cover of $X$, then we can define
a 1-gerbe on $X$ in terms of two pieces of data:
\begin{enumerate}
\item A principal $U(1)$ 
bundle ${\cal L}_{\alpha \beta}$ over
each $U_{\alpha \beta} = U_{\alpha} \cap U_{\beta}$
(subject to the convention ${\cal L}_{\beta \alpha} = {\cal L}_{\alpha
\beta}^{-1}$), such that ${\cal L}_{\alpha \beta} \otimes
{\cal L}_{\beta \gamma} \otimes {\cal L}_{\gamma \alpha}$
is trivializable on $U_{\alpha \beta \gamma}$
\item An explicit trivialization $\theta_{\alpha \beta \gamma}:
U_{\alpha \beta \gamma} \rightarrow U(1)$ of ${\cal L}_{\alpha \beta}
\otimes {\cal L}_{\beta \gamma} \otimes {\cal L}_{\gamma \alpha}$
on $U_{\alpha \beta \gamma}$
\end{enumerate}

Then, $\theta$ naturally defines a \v{C}ech 2-cochain, and it should
be clear that $\delta \theta = 1$, i.e., the canonical section
of the canonically trivial bundle defined by tensoring the obvious
twelve factors of principal $U(1)$ bundles.

Thus, $\theta$ defines a 2-cocycle, and it should be clear that 
this 2-cocycle is the same 2-cocycle defining the 1-gerbe in the
description in the previous subsection.

We should take a moment to clarify the precise relationship
between the construction above and 1-gerbes defined in terms
of sheaves of categories.  In the description of gerbes in terms
of sheaves of categories, one can define transition functions for
the gerbe with respect to a local trivialization, in precise analogy
to transition functions for bundles.  However, for 1-gerbes
the objects one associates to overlaps of open sets are not maps into
the group, but rather line bundles, subject to a constraint on
triple overlaps.  Put more directly, the description
given in the paragraphs above precisely describes transition functions
for a 1-gerbe.  The corresponding element of sheaf cohomology is merely
a characteristic class, classifying isomorphism classes.

Thus, the description of 1-gerbes given so far in this section
is technically a description of transition functions for 1-gerbes.
The reader may well wonder what is a 1-gerbe; the answer is,
a special kind of sheaf of categories.  Sheaves of categories and
related concepts have been banished to \cite{dt2}, but we shall
give a very quick flavor of the construction here.

Sometimes one speaks of ``objects'' of the 1-gerbe.
These are line bundles ${\cal L}_{\alpha}$ over 
open sets $U_{\alpha}$, such that ${\cal L}_{\alpha \beta} = 
{\cal L}_{\alpha} \otimes {\cal L}^{-1}_{\beta}$.
Objects exist locally on $X$, but in general will not exist
globally (unless the 1-gerbe is trivializable, meaning the associated
\v{C}ech 2-cocycle is trivial in cohomology).

In more formal treatments of gerbes, one often associates sheaves of
categories with gerbes\footnote{More precisely, there is a standard
method to associate sheaves of 1-categories and 2-categories to
1-gerbes and 2-gerbes, respectively.  The higher-degree gerbes
outlined in \cite{hitchin,dcthesis} presumably correspond to
sheaves of higher-degree multicategories, however the precise
definitions required have not been worked out, to our knowledge.}.
In this description, the ``objects'' mentioned above are precisely
objects of a category associated to some open set on $X$.
We shall not go into a detailed description of gerbes as sheaves
of categories in this section; see instead \cite{dt2}.

Now that we have discussed 1-gerbes, how are $n$-gerbes defined?
In principle an analogous description should hold true -- 
transition functions for an $n$-gerbe should consist of associating
an $(n-1)$-gerbe to each overlap, subject to constraints at
triple overlaps.  Although we are well-acquainted with more
intrinsic definitions of 1-gerbes \cite{dt2}, we have not worked
through higher $n$-gerbes in comparable detail, and so we hesitate
to say much more concerning transition functions for higher order gerbes. 
We hope to return to this elsewhere \cite{meinprog}.



\subsection{Connections on gerbes}

Now that we have stated the definition of an $n$-gerbe,
we shall define a connection on an $n$-gerbe, which is a straightforward
generalization of the notion of connection on a $C^{\infty}$ line
bundle.  

For simplicity, fix some good open cover $U_{\alpha}$ of $X$.
A connection on an $n$-gerbe is defined by a choice of 
$H \in \Omega^{n+2}(X)$ such that
$d H = 0$
(a closed $(n+2)$-form on $X$), and a choice of
$(B_{\alpha}) \in C^1( \Omega^{n+1} )$, that is,
a choice of smooth $(n+1)$-form on $U_{\alpha}$ for each $\alpha$,
such that on each $U_{\alpha}$,
$H |_{ U_{\alpha} } = d B_{\alpha}$, and such that on overlaps
$U_{\alpha} \cap U_{\beta}$, $B_{\alpha} - B_{\beta} = d A_{\alpha \beta}$,
where $A_{\alpha \beta}$ is a smooth $n$-form on $U_{\alpha \beta}$.
In general there will more more terms, of lower-degree-forms, filling
out an entire cocycle in the \v{C}ech-de Rham complex.

To be complete, we need to specify how the forms on various
open sets are related by the transition functions for the
$n$-gerbe.  For simplicity, consider a 1-gerbe.
Here, we have a globally-defined 3-form $H$,
a family of 2-forms $B_{\alpha}$, one for each open set $U_{\alpha}$,
a family of 1-forms $A_{\alpha \beta}$, one for each
intersection $U_{\alpha \beta} = U_{\alpha} \cap U_{\beta}$.
Recall transition functions for a 1-gerbe consist of
line bundles ${\cal L}_{\alpha \beta}$ associated to each
$U_{\alpha \beta}$; the 1-forms $A_{\alpha \beta}$ are precisely
connections\footnote{Note we are implicitly using the fact
that the $\{ U_{\alpha} \}$ form a good cover, so each
$U_{\alpha \beta}$ is contractible.} 
on the $U(1)$ bundles ${\cal L}_{\alpha \beta}$.
If $\theta_{\alpha \beta \gamma}$ denotes the specified trivialization
of ${\cal L}_{\alpha \beta} \otimes {\cal L}_{\beta \gamma}
\otimes {\cal L}_{\gamma \alpha}$ on $U_{\alpha \beta \gamma}$,
then we have
\begin{displaymath}
A_{\alpha \beta} \: + \: A_{\beta \gamma} \: + \: A_{\gamma \alpha} 
\: = \: d \, \mbox{ln } \theta_{\alpha \beta \gamma}
\end{displaymath}
Then, as per the description above, 
\begin{displaymath}
B_{\alpha} \: - \: B_{\beta} \: = \: d A_{\alpha \beta}
\end{displaymath}
on overlaps $U_{\alpha \beta}$,
and
\begin{displaymath}
H |_{U_{\alpha}} \: = \: d B_{\alpha}
\end{displaymath}
In principle, similar remarks hold for more general $n$-gerbes.

The reader should immediately recognize that a connection
on a $1$-gerbe is precisely the same thing as
a type II string theory $B$-field.  (The point that $B$ fields
and the relation $H = dB$ should really only be understood locally
has been made previously in the physics literature, albeit not usually in
terms of connections on gerbes; see for example \cite{freedetal}.) 
This relationship seems to be well understood in certain parts
of the field; we repeat it here simply to make this note more 
self-contained.  In general, the reader should recognize that
tensor field potentials appearing in type II supergravities often
look 
like connections on gerbes.

The reader should also notice that a connection on a $0$-gerbe
precisely coincides with the usual notion of connection
on a smooth line bundle.  To make this more clear,
change notation as follows:  change $H$ to $F$, and change
$B$ to $A$.  For a connection on a smooth line bundle,
locally we have the relation $F = d A$, but globally
this does not hold if $F$ is a nonzero element of $H^2(X, {\bf R})$.

In the special case that $F$ descends from an element of $H^2(X, {\bf Z})$
via the natural map $H^2(X, {\bf Z}) \rightarrow H^2(X, {\bf R})$,
then there exists a $C^{\infty}$ line bundle whose first Chern class
is represented by $F$.  (In particular, K\"ahler forms can be
interpreted as the first Chern classes of (holomorphic) line bundles
precisely when the K\"ahler form lies in the image of
$H^2(X, {\bf Z})$ in $H^2(X, {\bf R})$.)  Analogously, for an $n$-gerbe,
when the curvature $H$ descends from an element of $H^{n+2}(X, {\bf Z})$,
then there exists an $n$-gerbe whose characteristic class is defined by
$H$.

In fact, we have been slightly sloppy about certain details.
Suppose that $H^{n+2}(X, {\bf Z})$ contains torsion\footnote{In
the mathematical sense.}, that is, elements of finite order,
then if an $n$-gerbe-connection defines
an $n$-gerbe, it does not do so uniquely -- one will get several
$n$-gerbes, each of which has a characteristic class that descends
to the same element of $H^{n+2}(X, {\bf R})$.  Are these extra
degrees of freedom physically relevant -- in other words,
must there be an actual gerbe underlying these connections?

It is easy to see that an actual gerbe must
underlie such connections.  The point is that torsion elements of $H^{n+2}(X,
{\bf Z})$ contain physically relevant information, as was
noted in, for example, \cite{edbaryons}.

Given that $n$-gerbes can be loosely interpreted
as one generalization of line bundles,
the reader may wonder if there is some gerbe-analogue
of the holonomy of a flat $U(1)$ connection.
Indeed, it is possible to define the holonomy of
a flat $n$-gerbe-connection, though we shall not do so here.
Such holonomies have been observed in physics previously;
one example is \cite{ori1}.


As mentioned earlier, gerbes are often described in terms
of sheaves of categories.  There is a corresponding notion
of connection in such a description, which we have summarized
in \cite{dt2} and can also be found in \cite[section 5.3]{brylinski}.

\subsection{Gauge transformations of gerbes}    \label{gtbasic}

For principal $U(1)$-bundles there is a well-defined notion
of gauge transformation:  a gauge transformation is defined
by a map $f: X \rightarrow U(1)$.  What is the analogue
for $n$-gerbes?

We shall begin by describing gauge transformations
for 1-gerbes.  It can be shown that the analogue of a gauge
transformation for a 1-gerbe is given by a principal $U(1)$-bundle,
and a gauge transformation of a 1-gerbe with connection is defined by
a principal $U(1)$-bundle with connection.
(Strictly speaking, an equivalence class of principal $U(1)$-bundles,
but we shall defer such technicalities to later discussions.)
A rigorous derivation of this fact and related material is given
in \cite{dt2}.  We shall describe the implications of this
fact for connections on 1-gerbes, and for transition functions.

Intuitively, how does a principal $U(1)$-bundle act on a 1-gerbe?
Very roughly, the general idea is that given a bundle with connection
$A$, under a gauge transformation the $B$ field will
transform as $B \mapsto B + d A$.
(At the same level, we can see that only equivalence classes of
bundles with connection are relevant.  If $A$ and $A'$ differ
by a gauge transformation (of bundles), then $d A = d A'$,
and so they define the same action on $B$.)
In terms of sheaves of categories, a 1-gerbe is locally a category
of all principal $U(1)$-bundles, so given any one object in that
category, we can tensor with a principal $U(1)$-bundle to recover
another object.  This is essentially the action, in somewhat
loose language.

To properly describe how a principal $U(1)$-bundle acts on a 1-gerbe
requires understanding 1-gerbes in terms of sheaves of categories.
The reader might well ask, however, how a bundle acts on the transition
functions for a 1-gerbe?  
We described transition functions
for 1-gerbes by associating principal $U(1)$ bundles
to intersections $U_{\alpha \beta}$, together with an explicit
trivialization of \v{C}ech coboundaries.  
The reader should
(correctly) guess that a gauge transformation of a 1-gerbe,
at the level of transition functions,
should be a gauge transformation of the bundle on each coordinate overlap,
such that
the gauge transformations preserve the trivializations on triple
intersections.
In other words, a gauge transformation of a 1-gerbe should
be a set of maps $g_{\alpha \beta}: U_{\alpha \beta} \rightarrow U(1)$,
subject to the condition that $\delta g = 1$.
Put more simply still, a gauge transformation of a 1-gerbe is
precisely a principal $U(1)$-bundle.
Note that as expected by analogy with bundles, the transition functions are
invariant (the bundles on coordinate overlaps are unchanged by
gauge transformations).
Note that by analogy with bundles, one should expect a gauge
transformation to leave transition functions invariant -- and indeed,
our 1-gerbe gauge transformation does leave the transition functions
invariant, as a gauge transformation of each bundle is an automorphism
of the bundle.

How does a gauge transformation of a 1-gerbe act on a
connection on the 1-gerbe?  Principal bundles have well-defined
actions on gerbes; a unique specification of the action of
a principal $U(1)$-bundle, call it $P$, on a gerbe connection is equivalent
to a choice of connection on $P$.
Let $\{ h_{\alpha \beta}: U_{\alpha \beta} \rightarrow U(1) \}$ be
transition functions for $P$, and
$\{ A^{\alpha} \}$ a set of 1-forms on elements of the cover $\{ U_{\alpha} \}$
defining a connection on $P$.
Let $( B^{\alpha}, A^{\alpha \beta}, g_{\alpha \beta \gamma} )$
be a set of data defining a connection on a 1-gerbe.
Then under the action of $P$, this data transforms as follows:
\begin{eqnarray*}
B^{\alpha} & \mapsto & B'^{\alpha} \equiv B^{\alpha} + d A^{\alpha} \\
A^{\alpha \beta} & \mapsto & A'^{\alpha \beta} \equiv A^{\alpha \beta} +
d \, \mbox{ln} \, h_{\alpha \beta} \\
g_{\alpha \beta \gamma} & \mapsto & g_{\alpha \beta \gamma} + \delta h \\
& & = g_{\alpha \beta \gamma}
\end{eqnarray*}

More generally, the reader should correctly guess that a gauge
transformation of an $n$-gerbe is an $(n-1)$-gerbe.  We shall not
attempt to justify this statement here, however.

\subsection{Gerbes versus K theory}

It has recently been claimed that the Ramond-Ramond tensor
field potentials of type II theories can be understood in terms
of K theory, so we feel we should take a moment to justify our
assumption of a gerbe description of certain fields.

In this paper, when we discuss discrete torsion, we have in
mind the NS two-form field potential of type II theories,
and we implicitly assume that the other tensor field potentials
have vanishing vacuum expectation value, to assure that
the curvature of the $B$ field is a closed form.
In these circumstances, the $B$ field is well-described as a
connection on a gerbe.

However, in general terms the basic ideas presented in this
paper should also hold more generally.  At some level, the point
of this paper is that in any theory containing fields with gauge
invariances, specifying the orbifold group action on the base space
does not suffice to define the action of the orbifold group
on the theory, because the orbifold group action can always
be combined with gauge transformations.
To properly understand possible equivariant structures
(equivalently, orbifold group actions) on tensor field potentials
not described as gerbes involves certain technical distinctions,
but the basic point is the same.

\subsection{Why gerbes?}

So far we have presented gerbes as being a natural mathematical structure
for which many of the tensor field potentials 
of supergravities can be understood as
connections.  A doubting Thomas might argue, are gerbes really necessary?
After all, in supergravity theories, we only see the tensor fields
themselves; why not only speak of tensor fields on coordinate charts,
and forget about more abstract underlying structures?

We shall answer this question with another question:  why bundles?
Whenever one sees a vector field with the usual gauge invariances,
it is commonplace to associate it with some bundle.
One can ask, why?  In supergravity and gauge theories containing
vector fields, one does not see a bundle, only a set of vector fields
on coordinate charts.  Bundles (formulated as topological spaces)
describe auxiliary spaces -- fibers -- which are fibered over spacetime,
but these auxiliary structures are neither seen nor detected 
in physics.  There are no extra dimensions in the theory corresponding
to the fiber of a fiber bundle, so why work with bundles at all?
Since using bundles means invoking physically meaningless auxiliary
structures, why not just ignore bundles and only work with
vector fields on coordinate patches?

Part of the reason people speak of bundles and not just vector fields on
coordinate patches is that bundles give an insightful, elegant
way of thinking about vector fields on coordinate patches.
For example, recent discussions of brane charges in terms of
K-theory \cite{gregruben,edktheory} would have been far more
obscure if the notion of bundles was not commonly accepted.

Similarly, the notion of a gerbe gives an insightful and elegant
structure in which to understand many of the tensor field potentials appearing
in supergravity theories.
In principle, one could understand tensor fields without thinking
about gerbes, in the same way that one can understand vector fields
without thinking about bundles.  However, just as bundles give an insightful
and useful way to think about vector fields, so gerbes give an
insightful and useful way to think about tensor fields.

A slightly more subtle question that could be asked is the following.
In \cite{dt2}, we shall describe 1-gerbes in terms of sheaves of
categories; however, this description is not unique -- gerbes
can be described in several different ways.
If one should work with gerbes, which description is relevant?

A closely analogous problem arises in dealing with bundles.
A bundle has multiple descriptions -- as a topological space,
or a special kind of sheaf, for example.  Connections on bundles
can be described in terms of vector fields, or, in special circumstances,
as holomorphic structures on certain topological spaces.
The correct description one should use varies depending upon the application
and one's personal taste.  Similarly, which description of
gerbes is relevant varies depending upon both the application and  
personal inclination.

\section{Discrete torsion}      \label{dtmain}

In defining orbifolds, it is well-known that
the Riemannian space being orbifolded must have a well-defined
action of the orbifold group $\Gamma$.
However, after our discussion of gerbes, the reader should suspect
that something has been omitted from standard discussions of orbifolds
in string theory.
Namely, if we understand some of the tensor fields appearing in type II 
supergravities in terms of connections on gerbes, then we must
also specify the precise $\Gamma$-action on the gerbes.
This action need not be trivial, and (to our knowledge) has been
completely neglected in previous discussions of type II orbifolds.


We shall find that the set of actions of an orbifold group
$\Gamma$ on a 1-gerbe is naturally acted upon by a group
which includes $H^2(\Gamma, U(1))$, just as the set of
orbifold group actions on a principal $U(1)$-bundle with connection
is naturally acted upon by $H^1(\Gamma, U(1))$.
In special cases, such as trivial gerbes for example, there will be a canonical
orbifold group action, and in such cases we can identify the set of
orbifold group actions is identified with a group.
The possible equivariant structures (meaning, the possible orbifold
group actions) correspond to analogues of orbifold Wilson lines
for $B$-fields, in the same way that equivariant structures on a 
principal $U(1)$-bundle with connection correspond to orbifold
Wilson lines.

It is natural to speculate that the action of an orbifold
group $\Gamma$
on an $n$-gerbe
is described by a group including 
$H^{n+1}(\Gamma, U(1))$, in a fashion analogous to the above.  
This can be checked for 2-gerbes in the same
fashion as for 1-gerbes described in this paper, and we are
presently studying this issue \cite{meinprog}.  
For gerbes of higher degree, a precise understanding in terms
of sheaves of multicategories is not yet known, and so one
can only make somewhat more limited remarks \cite{meinprog}.

\subsection{Basics of discrete torsion}

Discrete torsion was originally discovered as an ambiguity in
the choice of phases of different twisted sector contributions
to partition functions of orbifold string theories.
The possible inequivalent choices of phases are counted
by elements of the group cohomology group\footnote{Where the
action of the group $\Gamma$ on the coefficients $U(1)$
is trivial.} $H^2(\Gamma, U(1))$,
where $\Gamma$ is the orbifold group.
Since its discovery, discrete torsion has been considered a rather mysterious
quantity.

Our description of discrete torsion essentially boils
down to the observation that discrete torsion is the
analogue of orbifold Wilson lines for 2-form-fields rather
than vectors.  Recall orbifold Wilson lines could be
described as a (discrete) ambiguity in lifting the orbifold
action on a space to a bundle with connection.
A similar ambiguity arises in lifting orbifold actions to
gerbes with connection, and this ambiguity is partially measured
by $H^2(\Gamma, U(1))$ \cite{dt2}.  More precisely, in general
the set of lifts of orbifold actions (in more technical
language, the set of equivariant structures on a (1-)gerbe
with connection) is (noncanonically) isomorphic
to a group which includes
$H^2(\Gamma, U(1))$, viewed as a set rather than a group.
In special cases, such as trivial gerbes, there exists a
canonical isomorphism.  

Just as for bundles, not all nontrivial gerbes admit
lifts of orbifold group actions.  We shall not attempt
to study conditions under which a nontrivial gerbe admits such a lift;
rather, we shall simply assume that lifts exist in all examples
in this paper.  

How does this description of discrete torsion as an analogue
of orbifold Wilson lines mesh with the original definition
in terms of distinct phases associated to twisted sectors of
string partition functions?
Recall there are factors
\begin{equation}    \label{phasefactor}
\exp\left( \, \int \, \phi^* B \, \right)
\end{equation}
in the partition function, contributing the holonomy of the $B$-field.
(We have used $\phi$ to denote the embedding map $\phi:
\Sigma \rightarrow X$ of the worldsheet into spacetime.)
Because of these holonomy factors, distinct lifts of the
orbifold action to the 1-gerbe with connection (i.e.,
distinct equivariant structures on the 1-gerbe with connection)
yield distinct phases in the twisted sectors of orbifold partition
functions -- we recover the original description of discrete torsion
\cite{vafadt}.  

Put another way, we do not just derive
some set of discrete degrees of freedom that happen
coincidentally to also be measured by $H^2(\Gamma, U(1))$; the discrete
degrees of freedom we recover necessarily describe discrete
torsion.  The passage from lifts of orbifold actions to phase
factors is provided by the partition function factors~(\ref{phasefactor}).

In passing, we should mention that the phase factors~(\ref{phasefactor})
were the original reason that discrete torsion, viewed as a 
set of phases of twisted sector contributions to partition functions,
was associated with $B$-fields at all \cite{vafadt}.
In some sense, our description of discrete torsion is a natural
outgrowth of some of the original ideas in \cite{vafadt}.  

In passing, we should also briefly speak to discrete torsion
on D-branes as discussed in \cite{md1,md2}.  In those references,
D-branes on orbifolds with discrete torsion were argued to be
described by specifying a projective representation of the orbifold
group on the bundle on the worldvolume.  We believe (although
we have not checked in total detail) that this can be 
derived from our description of discrete torsion, using
the interconnection between $B$-field backgrounds and bundles
on worldvolumes of D-branes, as recently described in 
\cite[section 6]{eddan}.

It is one thing to classify possible lifts of orbifold
actions to gerbes with connection;
it is quite another to describe precisely the 
characteristic classes and holonomies of the resulting
gerbe on the quotient space.  In principle, both could
be determined as for orbifold Wilson lines:
for a gerbe on a space $X$, with orbifold action $\Gamma$,
construct a gerbe on the space $E \Gamma \times_{\Gamma} X$,
such that the projection to $X / \Gamma$ yields the quotient gerbe,
analogously to the program pursued in \cite{freedvafa}.
We shall not pursue this program here.

Suppose the (discrete) orbifold group $\Gamma$ acting on $X$ 
acts freely, i.e., without fixed points.  In section~\ref{basicobw},
we studied moduli spaces of flat connections on quotient spaces,
in order to gain insight into orbifold Wilson lines.
In particular, we argued that (for bundles admitting flat connections)
orbifold $U(1)$ Wilson lines could be understood
directly in terms of extra elements of $\mbox{Hom}( \pi_1, U(1) )$
in the quotient space, for $\Gamma$ a freely-acting discrete group.
In other words, in this case orbifold Wilson lines were precisely
Wilson lines on the quotient space.  
We shall perform analogous calculations here.
(In the next few paragraphs we shall implicitly only consider flat $n$-gerbes
-- but only for the purposes of performing illuminating calculations.
We do not make such assumptions elsewhere.)

For gerbes, the interpretation is slightly more obscure.
First, note that in the case $\Gamma$ acts freely,
we have a principal $\Gamma$ bundle $\Gamma \rightarrow X
\rightarrow X/\Gamma$, so we can apply the long exact sequence
for homotopy to find that $\pi_n(X) = \pi_n(X/\Gamma)$
for $n > 1$ and $\Gamma$ discrete.
In other words, although the fundamental group of $X/\Gamma$
received a contribution from $\Gamma$, the higher homotopy
groups of $X/\Gamma$ are identical to those of $X$.
Thus, the higher-dimensional analogues of orbifold Wilson lines
(for flat $n$-gerbes)
can not correspond to extra elements of homotopy groups.
We shall find, rather, that the higher-dimensional analogues
correspond to extra elements of
\begin{displaymath}
\mbox{Hom}_{{\bf Z}}\left( \, H_n( X/\Gamma ), U(1) \, \right)
\end{displaymath}

We shall now describe the homology of the quotient
$X/\Gamma$.
One way to compute the homology
of the quotient space $X / \Gamma$ 
is as the limit of the Cartan-Leray spectral sequence
\cite[section VII.7]{brown}
\begin{equation}    \label{cartanleray}
E^2_{p,q} \: = \: H_p( \Gamma, H_q(X) )
\end{equation}
Note that the group homology appearing in the definition
above has the property that, in general, $\Gamma$ acts
nontrivially\footnote{An example should make this clear.
Let $X$ be the disjoint union of 2 identical disks,
and let $\Gamma$ be a ${\bf Z}_2$ exchanging the two disks.
Then $H_0(X) = {\bf Z}^2$, and $\Gamma$ exchanges the two ${\bf Z}$
factors, i.e., $\Gamma$ acts nontrivially on $H_0(X)$.} 
on the coefficients $H_q(X)$, even if $\Gamma$ acts
freely on $X$.

In special cases, the homology of $X/\Gamma$ can be computed
more directly.  For example, for any path-connected space $Y$,
for any $n > 1$ such that $\pi_r(Y) = 0$ for all $1 < r < n$,  
we have that
\cite[theorem II]{eilmac} the following sequence is exact:
\begin{displaymath}
0 \: \longrightarrow \: \pi_n(Y) \: \longrightarrow \:
H_n(Y) \: \longrightarrow \: H_n( \pi_1(Y), {\bf Z}) \:
\longrightarrow \: 0
\end{displaymath}
where the group homology $H_n( \pi_1(Y), {\bf Z})$
is defined by the group $\pi_1(Y)$ having trivial action
on the coefficients ${\bf Z}$.
Using the results above, we find that for path-connected $X$
such that $\pi_r(X) = 0$ for $1 < r < n$ for some $n > 1$,
the following sequence is exact:
\begin{displaymath}
0 \: \longrightarrow \: \pi_n(X) \: \longrightarrow \:
H_n(X / \Gamma) \: \longrightarrow \: 
H_n( \pi_1(X/\Gamma) , {\bf Z} ) \: \longrightarrow \: 0
\end{displaymath}
In the special case that $\pi_1(X) = 0$, we can rewrite this as
\begin{displaymath}
0 \: \longrightarrow \: \pi_n(X) \: \longrightarrow \:
H_n(X/\Gamma) \: \longrightarrow \: H_n(\Gamma, {\bf Z})
\: \longrightarrow \: 0
\end{displaymath}
Moreover, using the Hurewicz theorem, applying the
functor $\mbox{Hom}_{{\bf Z}}(-, U(1))$, and using
a relevant universal coefficient theorem, we can rewrite the
short exact sequence above as\footnote{This result has been
independently derived, using other methods, by P.~Aspinwall.}
\begin{displaymath}
0 \: \longrightarrow \: H^n(\Gamma, U(1)) \: \longrightarrow \:
\mbox{Hom}_{{\bf Z}}\left( H_n(X/\Gamma), U(1) \right) \:
\longrightarrow \: \mbox{Hom}_{{\bf Z}}\left( H_n(X), U(1) \right)
\: \longrightarrow \: 0
\end{displaymath}
(Technically we are also assuming that $X/\Gamma$ is a path-connected
space.)

From the calculation above we can extract two important lessons.
First, for $n=2$, we see that (in special cases), the holonomy
of a $B$-field (of a flat 1-gerbe) 
on the quotient $X/\Gamma$, as measured by
$\mbox{Hom}(H_2(X/\Gamma), U(1))$, differs from the possible
holonomies on the covering space by $H^2(\Gamma, U(1))$, and so
we can understand discrete torsion in such cases as being
precisely the extra contribution to $\mbox{Hom}(H_2(X), U(1))$
on the quotient.  (More generally, the precise relationship
between group cohomology and holonomies of $B$-fields is
described by the spectral sequence~(\ref{cartanleray}).)

The second, and more basic, lesson
we can extract from the calculation above is that
it is quite reasonable to believe that there exist analogues of
discrete torsion and orbifold Wilson lines for the higher-ranking
tensor fields appearing in supergravity theories, and that those
analogues of discrete torsion should be measured by 
higher-degree group cohomology $H^n(\Gamma, U(1))$.
We shall return to this point later.  In this paper, we only
derive\footnote{Our derivation in \cite{dt2} is not restricted to flat 1-gerbes;
the restriction to flat 1-gerbes in the previous few paragraphs
was for purposes of making illustrative calculations only.
We should also mention that our derivation in \cite{dt2}
does not assume $\Gamma$ is freely
acting, or that it is abelian -- our derivation holds equally
well regardless.} discrete torsion for $B$-fields, and in so doing find 
$H^2(\Gamma, U(1))$.  However, our general methods should apply
equally well to higher-ranking tensor fields, and it is extremely
tempting to conjecture that the analogue of discrete torsion for
an $n$-gerbe is measured by $H^{n+1}(\Gamma, U(1))$.

Before we go on to outline how discrete torsion can be derived,
we shall mention that in this paper, when speaking of an $n$-gerbe
on a space $X$, we shall assume that $X$ has (${\bf R}$) dimension
at least $n$.

\subsection{Equivariant gerbes}

In this section we shall try to give some intuitive understanding
of the classification of equivariant structures on 1-gerbes,
that is, the classification of lifts of the orbifold action
to 1-gerbes.
More precisely, we shall study what equivariant structures on 1-gerbes
mean at the level of transition functions for 1-gerbes.
We shall not be able to rigorously derive results on equivariant
gerbes in this fashion -- such derivations are instead given in \cite{dt2}. 
However, we hope that this approach should give the
reader some intuitive understanding of our results, without requiring
them to gain a detailed understanding of 1-gerbes in terms of stacks.

Let ${\cal C}$ denote a 1-gerbe on a space $X$, and let $\Gamma$
denote a group acting on $X$ by homeomorphisms.
Let $\{ U_{\alpha} \}$ be a ``good invariant'' cover of $X$ -- namely,
a cover such that each $U_{\alpha}$ is invariant under $\Gamma$
and each $U_{\alpha}$ is a disjoint union of contractible open sets.
(For example, we can often obtain such a cover as the inverse image
of a good cover on the quotient $X / \Gamma$.)
Note that a good invariant cover is not usually a good cover.

In order to define ${\cal C}$ at the level of transition functions for
the cover $\{ U_{\alpha} \}$, recall we need to specify a line bundle
${\cal L}_{\alpha \beta}$ on each overlap $U_{\alpha \beta} \equiv
U_{\alpha} \cap U_{\beta}$, and an explicit trivialization
$\theta_{\alpha \beta \gamma}: U_{\alpha \beta \gamma} \rightarrow U(1)$
of ${\cal L}_{\alpha \beta} \otimes {\cal L}_{\beta \gamma} \otimes
{\cal L}_{\gamma \alpha}$ on $U_{\alpha \beta \gamma}$.

Now, let us describe how one defines an equivariant structure
on the 1-gerbe ${\cal C}$ at the level of transition functions.
First, we need $g^* {\cal L}_{\alpha \beta} \cong {\cal L}_{\alpha
\beta}$ for all $g \in \Gamma$.
Let $\psi^g_{\alpha \beta}: {\cal L}_{\alpha \beta} \stackrel{ \sim }{
\longrightarrow } g^* {\cal L}_{\alpha \beta}$ denote a specific 
choice of isomorphism.  Since $\{ U_{\alpha} \}$ is a good invariant
cover of $X$, we can represent each $\psi^g_{\alpha \beta}$ by
a function $\nu^g_{\alpha \beta}: U_{\alpha \beta} \rightarrow U(1)$.

Note that the $\theta_{\alpha \beta \gamma}$ necessarily now obey
\begin{equation}    \label{gtheta}
g^* \theta_{\alpha \beta \gamma} \: ( \, = \,  \theta_{\alpha \beta
\gamma} \circ g \, ) \:  
 \: = \:  \theta_{\alpha \beta \gamma} \, \nu^g_{\alpha \beta} \, 
\nu^g_{\beta \gamma} \, \nu^g_{\gamma \alpha}
\end{equation}

Before going on, we should pause to derive an implication of 
equation~(\ref{gtheta}).  Let $\nu^g_{\alpha \beta}$ and
$\overline{\nu}^g_{\alpha \beta}$ denote a pair of maps
(partially) defining equivariant structures on ${\cal C}$.
Define 
\begin{displaymath}
\gamma^g_{\alpha \beta} \: \equiv \: \frac{ \nu^g_{\alpha \beta} }{
\overline{\nu}^g_{\alpha \beta} }
\end{displaymath}
then the $\gamma^g_{\alpha \beta}$ satisfy
\begin{displaymath}
\gamma^g_{\alpha \beta} \, \gamma^g_{\beta \gamma} \, 
\gamma^g_{\gamma \alpha} \: = \: 1
\end{displaymath}
for all $g \in \Gamma$, and so define transition functions for a 
bundle on $X$ we shall denote $T_g$.
Thus, even though we have not finished describing equivariant structures
on the 1-gerbe ${\cal C}$ at the level of transition functions,
we can already derive the fact that any two
equivariant structures will differ by,
among other things, a set of principal $U(1)$-bundles $T_g$,
one for each $g \in \Gamma$.

Before we can claim to have defined an equivariant structure on the
transition functions for ${\cal C}$, we need to fill in a few more
details.  In particular, how do the $\nu$ behave under composition 
of actions of elements of $\Gamma$?  We shall demand that for any
pair $g_1, g_2 \in \Gamma$,
\begin{equation}    \label{constr1}
(\nu^{g_2}_{\alpha \beta}) \, g_2^* (\nu^{g_1}_{\alpha \beta}) \: = \:
(\nu^{g_1 g_2}_{\alpha \beta}) \, h(g_1, g_2)_{\alpha} \, 
h(g_1, g_2)_{\beta}^{-1}
\end{equation}
for some functions $h(g_1, g_2)_{\alpha}: U_{\alpha} \rightarrow U(1)$.
We shall also demand that the functions $h(g_1, g_2)_{\alpha}$ satisfy
\begin{equation}   \label{constr2}
h(g_1, g_2)_{\alpha} \, h(g_1 g_2, g_3)_{\alpha} \: = \:
h(g_2, g_3)_{\alpha} \, h(g_1, g_2 g_3)_{\alpha} 
\end{equation}
These constraints probably seem relatively unnatural to the reader.
In our discussion of equivariant gerbes in terms of stacks,
we shall show how these constraints (or, rather, their more complete
versions for stacks) are quite natural.

We can attempt to rewrite equations~(\ref{constr1}) and (\ref{constr2})
somewhat more invariantly in terms of the line bundles
${\cal L}_{\alpha \beta}$ on overlaps $U_{\alpha \beta} = U_{\alpha}
\cap U_{\beta}$.  Recall that $\nu^g_{\alpha \beta}$ is the local
trivialization representation of the bundle morphism
$\psi^g_{\alpha \beta}: {\cal L}_{\alpha \beta} \rightarrow
g^* {\cal L}_{\alpha \beta}$,
then equation~(\ref{constr1}) states that the two bundle morphisms
\begin{displaymath}
(g_2^* \psi^{g_1}_{\alpha \beta}) \circ \psi^{g_2}_{\alpha \beta}:
\: {\cal L}_{\alpha \beta} \: \longrightarrow \:
(g_1 g_2)^* {\cal L}_{\alpha \beta}
\end{displaymath}
and
\begin{displaymath}
\psi^{g_1 g_2}_{\alpha \beta}: \: {\cal L}_{\alpha \beta} \:
\longrightarrow \: (g_1 g_2)^* {\cal L}_{\alpha \beta}
\end{displaymath}
are related by a gauge transformation on $(g_1 g_2)^* {\cal L}_{\alpha
\beta}$ defined by $h(g_1, g_2)_{\alpha} h(g_1, g_2)_{\beta}^{-1}$,
i.e.,
\begin{displaymath}
(g_2^* \psi^{g_1}_{\alpha \beta}) \circ \psi^{g_2}_{\alpha \beta}
\: = \: \kappa \circ \psi^{g_1 g_2}_{\alpha \beta}
\end{displaymath}
where $\kappa: (g_1 g_2)^* {\cal L}_{\alpha \beta} \rightarrow
(g_1 g_2)^* {\cal L}_{\alpha \beta}$ is the gauge transformation
defined by the function $h(g_1, g_2)_{\alpha} h(g_1, g_2)_{\beta}^{-1}$
on $U_{\alpha \beta}$.

Given two distinct equivariant structures on the same transition functions,
labelled by $\nu, \overline{\nu}$ and $h, \overline{h}$, if we define
functions
\begin{displaymath}
\omega(g_1, g_2)_{\alpha} \: \equiv \: \frac{ h(g_1, g_2)_{\alpha} }
{ \overline{h}(g_1, g_2)_{\alpha} }
\end{displaymath}
then from equation~(\ref{constr1}) we have the relation
\begin{equation}    \label{wdef1}
(\gamma^{g_2}_{\alpha \beta}) \, g_2^* (\gamma^{g_1}_{\alpha \beta}) \: = \:
(\gamma^{g_1 g_2}_{\alpha \beta}) \, \omega(g_1, g_2)_{\alpha} \,
\omega(g_1, g_2)_{\beta}^{-1}
\end{equation}
The functions $\omega(g_1, g_2)_{\alpha}$ define local trivialization
realizations of isomorphisms of principal
$U(1)$-bundles.  We denote these bundle isomorphisms by $\omega_{g_1, g_2}$,
and so we can rewrite equation~(\ref{wdef1}) more invariantly as
the definition of $\omega_{g_1, g_2}$:
\begin{displaymath}
\omega_{g_1, g_2}: \: T_{g_1 g_2} \: \stackrel{ \sim }{ \longrightarrow } \:
T_{g_2} \cdot g_2^* T_{g_1}
\end{displaymath}

Furthermore, from equation~(\ref{constr2}) we see that the bundles $T_g$ and
isomorphisms $\omega_{g_1, g_2}$ are further related by
\begin{equation}    \label{equivgerbediag}
\begin{array}{ccc}
T_{g_1 g_2 g_3} & \stackrel{ \omega_{g_1 g_2, g_3} }{ \longrightarrow } &
T_{g_3} \cdot g_3^* T_{g_1 g_2} \\
\makebox[0pt][r]{ $\scriptstyle{ \omega_{g_1, g_2 g_3} }$ } \downarrow
& & \downarrow \makebox[0pt][l]{ $\scriptstyle{ \omega_{g_1, g_2} }$ } \\
T_{g_2 g_3} \cdot (g_2 g_3)^* T_{g_1} & \stackrel{ \omega_{g_2, g_3} }{
\longrightarrow } & T_{g_3} \cdot g_3^*( T_{g_2} \cdot g_2^* T_{g_1} ) 
\end{array}
\end{equation}

So far we have argued that the difference between two equivariant
structures on a 1-gerbe is determined by the data $(T_g, 
\omega_{g_1, g_2})$, where the $\omega$ are required to make
diagram~(\ref{equivgerbediag}) commute.
However, it should also be said that the bundles $T_g$ are only
determined up to isomorphism.
Given a set of principal bundle isomorphisms $\kappa_g:
T_g \rightarrow T'_g$, we can replace the data $(T_g,
\omega_{g_1, g_2})$ by the data
$\left( T'_g, \kappa_{g_1 g_2} \circ \omega_{g_1, g_2}
\circ ( \kappa_{g_2} \cdot  g_2^* \kappa_{g_1} )^{-1} \right)$
and describe the same difference between equivariant structures.

\subsection{Equivariant gerbes with connection}

To properly derive the classification of equivariant gerbes
with connection at the level of transition functions is
rather messy and not very illuminating, so 
instead we shall settle for outlining the
main points.  (A complete derivation, in terms of gerbes as
stacks, can be found in \cite{dt2}, and a complete derivation at the level
of transition functions, as well as related information,
will appear in \cite{dt3}.)

In the previous section, we argued
that any two equivariant structures on a (1-)gerbe differ
by a set of principal $U(1)$ bundles $T_g$ ($g \in \Gamma$),
together with appropriate bundle isomorphisms $\omega_{g_1, g_2}$,
such that diagram~(\ref{equivgerbediag}) commutes,
modulo isomorphisms of bundles.

A gauge transformation of a gerbe with connection is defined
by a principal $U(1)$-bundle with connection, so the reader
should not be surprised to hear that the difference between
two equivariant structures on a 1-gerbe with connection
is defined by bundles $T_g$ with connection, together
with connection-preserving maps $\omega_{g_1, g_2}$ such that
diagram~(\ref{equivgerbediag}) commutes.
Also, the connections on the bundles $T_g$ are constrained to be flat.

Note that this structure is closely analogous to the discussion of
orbifold $U(1)$ Wilson lines.  In both cases, we find that the
difference between two equivariant structures is determined by
a set of gauge transformations, such that the gauge transformation
associated to the product $g_1 g_2$ is isomorphic to the product
of the gauge transformations associated to $g_1$ and $g_2$.
The constraint for bundles that the gauge transformations be constant
becomes the present constraint that the gerbe gauge transformations
defined by bundles with connection, must have a flat connection.

Just as before, the bundles $T_g$ are only defined up to equivalence.
We can replace any of the bundles $T_g$ with connection with
an isomorphic bundle with connection (changing $\omega_{g_1, g_2}$
appropriately), and describe the same difference between 
equivariant structures.

%
%

Where does $H^2(\Gamma, U(1))$ appear in this structure?
Assume for simplicity that $X$ is connected.
Take all the bundles $T_g$ to be topologically trivial,
with gauge-trivial connections.  Then, by replacing these bundles
with isomorphic bundles, we can assume without loss of generality
that each bundle $T_g$ is the canonical trivial bundle,
with identically zero connection.
The morphisms $\omega_{g_1, g_2}$ are now gauge transformations
of the canonical trivial bundle, and since
they must preserve
the connection, they must be constant maps.
From 
commutivity of diagram~(\ref{equivgerbediag}),
it is clear that any set of $\omega_{g_1, g_2}$ defines a cocycle
representative of an element of $H^2(\Gamma, U(1))$ (with
trivial action of $\Gamma$ on the coefficients $U(1)$), in the
inhomogeneous representation. 
Now, we still can act on any of the $T_g$ by constant
gauge transformations to get isomorphic equivariant
structures, and it is easy to see that these define
group coboundaries.

Now, in general not every set of data $(T_g, \omega_{g_1, g_2})$
corresponds to topologically-trivial $T_g$ with gauge-trivial
connection -- the $T_g$ are only constrained to be flat,
so it is not difficult to find new degrees of freedom that
do not correspond to elements of $H^2(\Gamma, U(1))$.
We shall discuss these further in \cite{dt3}.

In special cases, such as trivial gerbes, there exist
canonical trivial equivariant structures, and so 
elements of the group $H^2(\Gamma, U(1))$ can be identified with
(some of) the equivariant structures.
More generally,
the identification of (some) equivariant structures with
$H^2(\Gamma, U(1))$ is not canonical\footnote{Technically,
in general the set of equivariant structures on a gerbe
with connection is merely a torsor.}.


\subsection{Analogues of discrete torsion}

So far in this paper we have outlined how orbifold Wilson lines
and discrete torsion both appear as an ambiguity in the
choice of orbifold group action on some tensor field potential.
Although we have only concerned ourselves with vector fields and
NS-NS $B$ fields, in principle analogous ambiguity exists for
every tensor field potential appearing in string theory.

Put another way, in general whenever one has a theory containing
fields with gauge invariances, specifying an orbifold group action
on the base space does not suffice to define the orbifold group
action on the fields of the theory, as one can combine any orbifold
group action with gauge transformations.

For example, the other RR tensor field potentials of type II
theories should also have analogues of discrete torsion,
given as the ambiguity in the choice of orbifold group
action on the fields.  It has recently been pointed out that
these fields should be understood in terms of K-theory,
so given some Cheeger-Simons-type description of K-theory,
one should be able to calculate the analogues of discrete
torsion for these other fields.

What might analogues of discrete torsion be for other
tensor field potentials?  There is an obvious conjecture.
For vector fields, we found that the set of equivariant structures
is a torsor under the group $H^1(\Gamma, U(1))$.
For $B$ fields, we found that the set of equivariant structures
is a torsor under a group which includes $H^2(\Gamma, U(1))$.
Therefore, for a rank $p$ tensor field potential,
it is tempting to conjecture that the set of equivariant
structures is a torsor under some group which includes
$H^p(\Gamma, U(1))$.  We are presently studying this
matter \cite{meinprog}.

%

The reader might well ask how such degrees of freedom could be
seen in perturbative string theory.  Orbifold Wilson lines and
discrete torsion both crop up unavoidably; but how could one
turn on analogues for RR fields?  The answer surely lies in
the description of RR field backgrounds in perturbative string theory.
Judging from the results in,
for example, \cite{bvw,berenleigh,janshat,berketal,berk}, 
it seems reasonable to assume that 
one can understand Ramond-Ramond backgrounds
in conformal field theory after coupling to the superconformal ghosts,
so in principle analogues
of discrete torsion for RR fields in conformal field theory
might emerge when considering orbifolds of such backgrounds.   
Unfortunately, it might also be true that the RR analogues of discrete
torsion are simply not visible in string perturbation theory.

It is quite possible that there may also be certain analogues
of modular invariance conditions for these analogues of
discrete torsion.  We have only discussed gerbes in isolation,
whereas in type II theories, the gerbes interact with one another
(and so cannot really be understood as gerbes).
It is quite conceivable that, in order for any given orbifold
to define a symmetry of the full physical theory, there are
nontrivial constraints among analogues of discrete torsion for
various gerbes.  We have nothing particularly concrete to say
on this matter, though we hope to return to it in \cite{meinprog}.

It is not clear, however, whether all analogues of modular invariance
conditions can be described in this fashion.
For example, in \cite{fhsv} it was argued that there existed
a constraint on orbifold Wilson lines associated to the IIA RR 1-form,
arising nonperturbatively.  (We are referring to the
so-called ``black hole level matching'' of that reference.)
Unfortunately we are not able to address the existence and interpretation
of such constraints.

\section{Conclusions}

In this paper we have given a geometric description of discrete
torsion, as a precise analogue of orbifold Wilson lines. 
Put another way, we have described discrete torsion as
``orbifold Wilson surfaces.''
After giving a mathematically precise discussion of orbifold
Wilson lines, we outlined how the classification of orbifold
Wilson lines (as equivariant structures on bundles with connection)
could be extended to discrete torsion (as equivariant structures
on 1-gerbes with connection).  Although we outlined how this
result on discrete torsion was proven, we have deferred a
rigorous examination to \cite{dt2}.


\section{Acknowledgements}

We would like to thank P.~Aspinwall, D.~Freed,
A.~Knutson, D.~Morrison, and R.~Plesser
for useful conversations, and E.~Silverstein for pointing
out reference~\cite{freedvafa}.  
We would also like to thank J.~Stasheff for extremely extensive comments
on the first version of this paper, and we would like to apologize
for taking so long to implement his suggestions.

\appendix

\section{Review of group cohomology}

For a complete technical overview of group cohomology,
the standard reference is \cite{brown}.  For much shorter
and more accessible accounts, we recommend \cite[section IV.4]{amp2}
and \cite{dw}.

Let $G$ and $M$ be groups, $M$ abelian, with a (possibly trivial) action
of $G$ on $M$ by group automorphisms.
We shall assume that
the action of $G$ commutes with the group operation of $M$ on itself.

Define $C^n(G, M)$ to be the set of all maps
\begin{displaymath}
\epsilon: G \times \cdots \times G = G^{n+1} \rightarrow M
\end{displaymath}
such that $\epsilon( g g_0, g g_1, \cdots, g g_n) = g \epsilon( g_0,
g_1, \cdots, g_n)$ for all $g, g_i \in G$.
(This representation of the cochains is known as a homogeneous representation,
because of the obvious analogy with projective spaces.)

Define a coboundary operator $\delta: C^n(G, M) \rightarrow C^{n+1}(G,M)$ by
\begin{displaymath}
( \delta \epsilon )( g_0, \cdots, g_{n+1}) \: = \: \sum_{k=0}^{n+1}
\, (-)^k
\epsilon( g_0, \cdots, \hat{ g_k }, \cdots, g_{n+1} ) 
\end{displaymath}
Note that $\delta^2 \epsilon = 1$.

Define $Z^n(G,M)$ to be the set of cocycles, that is,
$\epsilon \in \mbox{ker } \delta \subset C^n(G, M)$.
Define $B^n(G,M)$ to be the set of coboundaries, that is,
$\epsilon \in \mbox{im } \delta \subset C^n(G, M)$.
Then define the group cohomology to be
$H^n(G,M) = Z^n(G,M) / B^n(G,M)$.

There is an alternative presentation of group cohomology,
which can be defined as follows.
Given a cochain $\epsilon \in C^n(G,M)$,
which is to say, a map $G^{n+1} \rightarrow M$,
define a map $\tilde{ \epsilon }: G^n \rightarrow M$ as,
\begin{displaymath}
\tilde{ \epsilon }(g_1, g_2, \cdots, g_n ) \: = \:
\epsilon( e, g_1, g_1 g_2, g_1 g_2 g_3, \cdots, g_1 g_2 \cdots g_n )
\end{displaymath}
This is known as an inhomogeneous representation, that is, these
are called inhomogeneous cochains.
It is then easy to demonstrate that
\begin{eqnarray*}
(\delta \tilde{ \epsilon })(g_1, g_2, \cdots, g_{n+1}) & = &
g_1 \tilde{ \epsilon }(g_2, \cdots, g_{n+1})  \\
& & + \: \sum_{k=1}^{n} (-)^k \tilde{ \epsilon }( g_1, g_2, \cdots,
g_k g_{k+1}, \cdots, g_{n+1} ) \\
& & + \: (-)^{n+1} \tilde{ \epsilon }(g_1, g_2, \cdots, g_n) 
\end{eqnarray*}

In the group cohomology appearing in this paper, and to our knowledge in 
the physics literature to date\footnote{For example, experts should note that
it is this latter, inhomogeneous form, restricted to the special case
that the action of $G$ on $M$ is trivial, which appears in \cite{dw}.}, 
we always
assume that the action of the group on the coefficients is trivial.

When the action of $G$ on $M$ is assumed trivial, if $\epsilon: G \rightarrow
M$
is a homogeneous 0-cochain, then it is easy to check that $\epsilon$
is constant.  From the definitions of coboundaries for homogeneous
and inhomogeneous cochains, it is easy to derive that the
associated inhomogeneous 0-cochain $\tilde{\epsilon}$ must always
be the identity of $M$.  To repeat, if $\tilde{\epsilon}$ is
an inhomogeneous 0-cochain, then $\tilde{\epsilon} = 1 \in M$. 

As a consequence, for trivial action of $G$ on $M$, we have
that $H^1(G, M) = Z^1(G, M)$, that is, $H^1(G, M)$ is precisely 
the set of group homomorphisms $G \rightarrow M$.

For group 2-cochains (defined with trivial group action on the
coefficients), there is a gauge choice that is often used.
From manipulating the group 2-cocycle condition (in inhomogeneous
representation), it is easy to check that
$\tilde{\epsilon}(1,g) = \tilde{\epsilon}(g,1) = \tilde{\epsilon}(1,1)$
for any $g$.  For convenience,
one often sets $\tilde{\epsilon}(1,1) = 1$
(just pick a group coboundary conveniently).  Then, in this gauge,
$\tilde{\epsilon}(1,g) = \tilde{\epsilon}(g,1) = 1$ for all $g$.
One is still free to add any group coboundary in this gauge,
modulo the constraint that if $\mu(g)$ defines a group
coboundary, one needs $\mu(1) = 1$ in order to stay in the gauge.

In passing, we shall mention that more formally,
for any $G$-module $M$, we can define group cohomology
as
\begin{displaymath}
H^n(G,M) \: \equiv \: \mbox{Ext}^n_{ {\bf Z}[G] }( {\bf Z}, M )
\end{displaymath}
where ${\bf Z}[G]$ is the free ${\bf Z}$-module generated by the 
elements of 
$G$.  In other words, any element of ${\bf Z}[G]$ can be written
uniquely in the form
\begin{displaymath}
\sum_{g \in G} \: a(g) g
\end{displaymath}
where $a(g) \in {\bf Z}$.
This definition of group cohomology does not make any assumptions concerning
the nature of the $G$-action on $M$.

In addition to group cohomology, one can also define group
homology in a very similar manner, though we shall not do so here.
For the case of group homology and cohomology defined
by groups with trivial actions on the coefficients,
there exist precise analogues of the usual universal coefficient
theorems for homology and cohomology \cite[exercise III.1.3]{brown}.
There is also a K\"unneth formula \cite[section V.5]{brown}.

For reference, we shall now list some commonly used
group homology and cohomology groups.
First, the homology groups $H_i({\bf Z}_n, {\bf Z})$,
where the group ${\bf Z}_n$ acts trivially on the coefficients
${\bf Z}$, are given by
\begin{displaymath}
H_i( {\bf Z}_n, {\bf Z}) \: = \: \left\{
\begin{array}{ll}
{\bf Z} & i = 0 \\
{\bf Z}_n & i \mbox{ odd} \\
0 & i \mbox{ even}, \: i > 0
\end{array}
\right.
\end{displaymath}
The cohomology groups $H^i({\bf Z}_n, U(1))$,
where the group ${\bf Z}_n$ acts trivially on the
coefficients $U(1)$, are given by
\begin{displaymath}
H^i({\bf Z}_n, U(1)) \: = \: \left\{
\begin{array}{ll}
U(1) & i = 0 \\
{\bf Z}_n & i \mbox{ odd} \\
0 & i \mbox{ even}, \: i > 0
\end{array}
\right.
\end{displaymath}

From the K\"unneth formula \cite[section V.5]{brown},
we find that
the homology groups $H_i({\bf Z}_n \times {\bf Z}_m, {\bf Z})$,
where the group acts trivially on the coefficients ${\bf Z}$,
are given by
\begin{displaymath}
H_i( {\bf Z}_n \times {\bf Z}_m, {\bf Z} ) \: = \: \left\{
\begin{array}{ll}
{\bf Z} & i = 0 \\
{\bf Z}_n \oplus {\bf Z}_m \oplus \bigoplus_{(i-1)/2} \mbox{Tor}_1^{{\bf Z}}
( {\bf Z}_n, {\bf Z}_m ) & i \mbox{ odd} \\
\bigoplus_{i/2} \left( {\bf Z}_n \otimes_{{\bf Z}} {\bf Z}_m \right)
& i \mbox{ even}, \: i > 0 
\end{array}
\right.
\end{displaymath}
In other words,
\begin{eqnarray*}
H_0( {\bf Z}_n \times {\bf Z}_m, {\bf Z}) & = & {\bf Z} \\
H_1( {\bf Z}_n \times {\bf Z}_m, {\bf Z}) & = & {\bf Z}_n \oplus
{\bf Z}_m \\
H_2( {\bf Z}_n \times {\bf Z}_m, {\bf Z}) & = & 
\left( {\bf Z}_n \otimes_{{\bf Z}} {\bf Z}_m \right) \\
H_3( {\bf Z}_n \times {\bf Z}_m, {\bf Z}) & = &
{\bf Z}_n \oplus {\bf Z}_m \oplus \mbox{Tor}_1^{{\bf Z}}( {\bf Z}_n,
{\bf Z}_m ) \\
H_4( {\bf Z}_n \times {\bf Z}_m, {\bf Z}) & = &
\left( {\bf Z}_n \otimes_{{\bf Z}} {\bf Z}_m \right) \oplus
\left( {\bf Z}_n \otimes_{{\bf Z}} {\bf Z}_m \right)
\end{eqnarray*}
and so forth.

Using the identities
\begin{eqnarray*}
{\bf Z}_n \otimes_{{\bf Z}} {\bf Z}_m & = & {\bf Z}_{ {\it gcd}(n,m) } \\
\mbox{Tor}_1^{{\bf Z}}( {\bf Z}_n, {\bf Z}_m ) & = &
{\bf Z}_{ {\it gcd}(n,m) }
\end{eqnarray*}
and the appropriate universal coefficient theorem, one can compute
the cohomology groups $H^i({\bf Z}_n \times {\bf Z}_m, U(1))$,
where the group ${\bf Z}_n \times {\bf Z}_m$ is assumed to act
trivially on the coefficients $U(1)$:
\begin{displaymath}
H^i({\bf Z}_n \times {\bf Z}_m, U(1)) \: = \:
\left\{ \begin{array}{ll}
U(1) & i = 0 \\
{\bf Z}_n \oplus {\bf Z}_m \oplus \bigoplus_{ (i-1)/2 } {\bf Z}_{
{\it gcd}(n,m) } & i \mbox{ odd} \\
\bigoplus_{i/2} {\bf Z}_{ {\it gcd}(n,m) } & i \mbox{ even}, \: i > 0
\end{array}
\right.
\end{displaymath}
In other words,
\begin{eqnarray*}
H^0({\bf Z}_n \times {\bf Z}_m, U(1)) & = & U(1) \\
H^1({\bf Z}_n \times {\bf Z}_m, U(1)) & = & {\bf Z}_n \oplus {\bf Z}_m \\
H^2({\bf Z}_n \times {\bf Z}_m, U(1)) & = & {\bf Z}_{ {\it gcd}(n,m) } \\
H^3({\bf Z}_n \times {\bf Z}_m, U(1)) & = & {\bf Z}_n \oplus {\bf Z}_m
\oplus {\bf Z}_{ {\it gcd}(n,m) } \\
H^4({\bf Z}_n \times {\bf Z}_m, U(1)) & = & {\bf Z}_{ {\it gcd}(n,m) }
\oplus {\bf Z}_{ {\it gcd}(n,m) } 
\end{eqnarray*}
and so forth.  Note that we have used the notation
$\times$ and $\oplus$ in this subsection interchangeably.

\end{document}